\RequirePackage{fix-cm}
\documentclass[twocolumn,epjc3]{svjour3}
\smartqed 
\RequirePackage{graphicx}
\usepackage{amsmath}
\usepackage{epstopdf}

\usepackage{xspace}

\usepackage[english]{babel}

\usepackage{amsfonts}
\usepackage{amssymb}
\usepackage{dsfont}
\usepackage[mathscr]{eucal}
\usepackage{braket}
\usepackage{cite}
\usepackage{titleref}
\usepackage{varioref}
\usepackage{enumitem}
\usepackage{mathtools}
\usepackage{tabularx}

\usepackage{epsfig}

\usepackage[usenames,dvipsnames]{xcolor}
\usepackage{soul}

\newcommand{\rmd}{\ensuremath{\mathrm{d}}}
\newcommand{\rme}{\ensuremath{\mathrm{e}}}
\newcommand{\rmi}{\ensuremath{\mathrm{i}}}

\DeclareMathOperator{\diag}{diag}

\newcolumntype{L}{>{\raggedright\arraybackslash}X}

\journalname{Eur. Phys. J. C}
\begin{document}

\title{Evolution of confined quantum scalar fields in curved spacetime. Part~II
}
\subtitle{Spacetimes with moving boundaries in any synchronous gauge}


\author{Luis C.\ Barbado\thanksref{e1,addr1,addr2}
\and
Ana L.\ B\'aez-Camargo\thanksref{e2,addr1}
\and
Ivette Fuentes\thanksref{e3,addr1,addr3,addr4}
}

\thankstext{e1}{e-mail: luis.cortes.barbado@univie.ac.at}
\thankstext{e2}{e-mail: lucia.baez@univie.ac.at}
\thankstext{e3}{e-mail: I.Fuentes-Guridi@soton.ac.uk}


\institute{Quantenoptik, Quantennanophysik und Quanteninformation, Fakult\"at f\"ur Physik, Universit\"at Wien, Boltzmanngasse 5, 1090 Wien, Austria \label{addr1}
\and
Institut f\"ur Quantenoptik und Quanteninformation, \"Osterreichische Akademie der Wissenschaften, Boltzmanngasse 3, 1090 Wien, Austria \label{addr2}
\and
School of Mathematical Sciences, University of Nottingham, University Park, Nottingham NG7 2RD, United Kingdom \label{addr3}
\and
School of Physics and Astronomy, University of Southampton, Southampton SO17 1BJ, United Kingdom \label{addr4}
}

\date{Received: date / Accepted: date}

\maketitle

\begin{abstract}
We develop a method for computing the Bogoliubov transformation experienced by a confined quantum scalar field in a globally hyperbolic spacetime, due to the changes in the geometry and/or the confining boundaries. The method constructs a basis of solutions to the Klein-Gordon equation associated to each compact Cauchy hypersurface of constant time. It then provides a differential equation for the linear transformation between bases at different times. The transformation can be interpreted physically as a Bogoliubov transformation when it connects two regions in which a time symmetry allows for a Fock quantisation. This second article on the method is dedicated to spacetimes with timelike boundaries that do not remain static in any synchronous gauge. The method proves especially useful in the regime of small perturbations, where it allows one to easily make quantitative predictions on the amplitude of the resonances of the field. Therefore, it provides a crucial tool in the growing research area of confined quantum fields in table-top experiments. We prove this utility by addressing two problems in the perturbative regime: Dynamical Casimir Effect and gravitational wave resonance. We reproduce many previous results on these phenomena and find novel results in an unified way. Possible extensions of the method are indicated. We expect that our method will become standard in quantum field theory for confined fields.



\end{abstract}

\section{Introduction} \label{intro}

Quantum field theory in curved spacetime studies the evolution of quantum fields which propagate in a classical general relativistic background geometry. Beyond its core mathematical construction (see e.g.\ \cite{Haag:1963dh, BD1984, wald1994quantum, parker_toms_2009}), the theory has been successful in approaching different concrete problems, such as Hawking and Unruh radiations \cite{Hawking1975, Unruh1976} or cosmological particle creation~\cite{Parker1968, Parker1969, parker_toms_2009}. This has required the development of different mathematical techniques and simplifications adapted to each specific problem, which allow for quantitative theoretical predictions. A family of problems of especial interest are quantum fields confined in cavities and under the effect of small changes in the background geometry or the non-inertial motion of the cavity boundaries. The theoretical predictions on these problems may be tested experimentally in the near future~\cite{Howl2018, Ahmadi2014, Tian2015relativistic, Wilson2011observation}, thanks to the great improvement of the precision of quantum measurements in table-top experiments. Consequently, new mathematical techniques are necessary to address these problems and make quantitative predictions, which can then be contrasted with the experimental results.

In the preceding article \cite{Barbado_2020}, which we shall call ``Part~I'', we constructed a method for computing the evolution of a confined quantum scalar field in a globally hyperbolic spacetime, by means of a time-dependent Bogoliubov transformation. The method proved especially useful for addressing the kind of problems just mentioned, related to confined quantum fields undergoing small perturbations, although it is of general applicability (under some minor assumptions). However, the mathematical construction of Part~I only allowed to approach spacetimes without boundaries or with static boundaries in some synchronous gauge.

In this second article we extend the method to spacetimes with timelike boundaries which do not remain static in any synchronous gauge. The essence of the procedure remains the same, but we require a more involved mathematical construction than the one undertaken in Part~I. In particular, we need a specific treatment for each different boundary condition that we may impose to the field. The method is based on the foliation of the spacetime in compact spacelike Cauchy hypersurfaces using a time coordinate. The core idea is to construct a basis of modes naturally associated to each hypersurface, and then provide a differential equation in time for the linear transformation between the modes associated to two hypersurfaces at different times. This way, the evolution of the field in time is not obtained by solving the Klein-Gordon equation, but rather by solving a differential equation for a time-dependent linear transformation between the bases. Such linear transformation can be interpreted physically as a Bogoliubov transformation when it relates regions in which the time symmetry allows for a Fock quantisation in terms of particles associated to the corresponding bases of modes.

The conception of transferring the time evolution from the mode functions to the Bogoliubov transformation appears for the first time in the pioneer work by Parker \cite{Parker1969}. This idea of a time-dependent Bogoliubov transformation has since then been developed specifically for other concrete problems (see e.g.\ \cite{woodhouse1976particle, starobinsky1977rate, fulling1979remarks, birrell1980massive, ford1987gravitational, glenz2009study}). Our work is therefore a generalisation (for confined fields) of the previous specific results. Operationally, it is mostly inspired by the construction in~\cite{Jorma2013} for periodically accelerated cavities.

As in Part~I, the method is of general applicability (with minor assumptions), but proves especially useful in the regime of small perturbations, since it provides very simple recipes for computing the resonance spectrum and sensibility of the field to a given perturbation of the background metric or the boundary conditions. We show with concrete examples that, in the small perturbations regime, with this unique method it is possible to easily solve different problems, each of which has so far required its own specific (and way more involved) treatment. Moreover, we easily handle a so far unsolved problem, namely that of a quantum field inside a three-dimensional rigid cavity and perturbed by a gravitational wave. We manage to explain it physically as a combination of the direct effect of the gravitational wave on the field plus a Dynamical Casimir Effect.

The contribution of the method to the understanding of quantum fields in curved spacetime is threefold. First, as we just mentioned, its direct application to concrete problems within its range allows to easily solve many important problems of physical interest. Second, the general structure of the method is very likely to be extensible (with the necessary adaptations) to other scenarios, such as other quantum fields, boundary conditions or metric gauges \cite{dewitt1975quantum, dalvit2006dynamical, friis2013scalar, Barbero2019, dodonov2020fifty, benito2020}. And third, the mathematical time-dependent linear transformation obtained may be given a physical interpretation beyond the one in terms of particle quantisation considered here; for example, in relation to adiabatic expansions~\cite{BD1984, fulling_1989, junker2002adiabatic} or to approaches to quantum field theory in curved spacetime based on field-related quantities~\cite{Haag:1963dh, ashtekar1975quantum, wald1994quantum}.

The article is organised as follows. In Sect.~\ref{prelim} we state the general physical problem for which we construct the method, introducing the background metric, the field theory and the different assumptions that we consider; and also define three important mathematical objects that we use. The nuclear part of the article is Sect.~\ref{nuclear}. In this section we construct the basis of modes associated to each hypersurface of the foliation of the spacetime, and formally compute the time-dependent linear transformation between the modes of two different hypersurfaces. We give a differential equation and a formal solution for it, which constitute one of the two main results of the work. We also discuss the physical meaning of both the modes and the transformation. In Sect.~\ref{res} we consider the particularly important case of small perturbations and resonances, obtaining especially simple recipes for its solution, which constitute the other main result of the work. We apply the recipes to the Dynamical Casimir Effect and the gravitational wave perturbation problems. Finally, in Sect.~\ref{conclu} we present the summary and conclusions. In addition, in \ref{why_part_II} we explain why a different treatment as that of Part~I is needed in the case of ``moving'' boundaries. In \ref{props_bases} we prove that the properties we assign to the sets of modes that we build are fulfilled. In \ref{great_proof} we provide the detailed computation of the differential equation provided in Sect.~\ref{nuclear}. In \ref{simplif} we derive the expressions given in Sect.~\ref{res}. In \ref{proof_J} and \ref{proof_geom} we provide the derivation of auxiliary expressions used in \ref{simplif}. \ref{dirichlet_app} is dedicated to the case of Dirichlet vanishing boundary conditions. In \ref{no_zeros} we prove a necessary proposition about certain sets of eigenvalues. In \ref{recipes} we provide for convenience a summary of the useful formulae for the application of the method.\footnote{As mentioned, this article introduces a more involved mathematical construction, as compared to Part~I. Therefore, although the article is self-contained, it is mostly devoted to the development of such construction. The underlying ideas and the physical picture behind the method, which are analogous in both parts, are thus explained in more detail in Part~I.}

\section{Preliminaries}\label{prelim}

\subsection{Statement of the problem}\label{statement}

We consider a globally hyperbolic spacetime~$(M, g)$ of dimension~$N+1$ with timelike boundary~$\partial M$~\cite{Ake:2018dzz}. In this geometry we introduce a scalar field $\Phi$ satisfying the Klein-Gordon equation
\begin{equation}
g^{\mu \nu} \nabla_\mu \nabla_\nu \Phi - m^2 \Phi - \xi R \Phi = 0;
\label{klein-gordon}
\end{equation}
where~$m \geq 0$ is the rest mass of the field, $g^{\mu \nu}$ is the spacetime metric, $R$ its scalar curvature and~$\xi \in \mathds{R}$ is a coupling constant (we use natural units~$\hbar = c = 1$).

We impose one of the following two boundary conditions to the field:

\begin{enumerate}

	\item[a)] Dirichlet vanishing boundary conditions
\begin{equation}
\Phi(t, \vec{x}) = 0, \quad (t, \vec{x}) \in \partial M.
\label{dirichlet}
\end{equation}

	\item[b)] Neumann vanishing boundary conditions
\begin{equation}
n^\mu \nabla_\mu \Phi(t, \vec{x}) = 0, \quad (t,\vec{x}) \in \partial M;
\label{neumann_cov}
\end{equation}
where~$n^\mu(t, \vec{x})$ is the normal vector to~$\partial M$.

\end{enumerate}
We treat explicitly Dirichlet and Neumann vanishing boundary conditions since they are arguably the most common ones in physical problems. However, we do not discard that a specific treatment for other boundary conditions is also possible. The treatment of Dirichlet boundary conditions~(\ref{dirichlet}) requires a subtle reformulation of the boundary conditions themselves, which nonetheless does not modify the physical problem being addressed. Due to the need of a specific discussion, we leave Dirichlet boundary conditions for \ref{dirichlet_app}. Therefore, from here on we consider only Neumann boundary conditions~(\ref{neumann_cov}) (except for \ref{dirichlet_app} or unless otherwise stated).

Thanks to the global hyperbolicity, it is always possible to construct a Cauchy temporal function~$t$ in the full spacetime~\cite{Ake:2018dzz}. This provides a foliation in Cauchy hypersurfaces~$\Sigma_t$ of constant time. We introduce the Klein-Gordon inner product between two solutions of~(\ref{klein-gordon}), given by
\begin{multline}
\langle \Phi', \Phi \rangle := \\
- \rmi \int_{\Sigma_{\tilde{t}}} \rmd V_{\tilde{t}}\ \left[ \Phi'(\tilde{t}) \left. \partial_t \Phi(t)^* \right|_{t=\tilde{t}} - \Phi(\tilde{t})^* \left. \partial_t \Phi'(t) \right|_{t=\tilde{t}} \right];
\label{scalar_product}
\end{multline}
which, for convenience, we already evaluated at a given Cauchy hypersurface~$\Sigma_{\tilde{t}}$, with~$\rmd V_{\tilde{t}}$ being its volume element. Under the boundary conditions~(\ref{dirichlet}) or~(\ref{neumann_cov}) this inner product is independent of~$\Sigma_{\tilde{t}}$.

Finally, we introduce the three conditions on the Cauchy hypersurfaces and the temporal function that we need to ensure the applicability of the method. These conditions are:

\begin{enumerate}
	\item[A.]
		The Cauchy hypersurfaces~$\Sigma_t$ must be compact.

	\item[B.]
		For any Cauchy hypersurface~$\Sigma_t$, the Cauchy problem for the Klein-Gordon equation~(\ref{klein-gordon}) must be well-posed; that is, given as initial conditions the value of the field and of its first derivative with respect to~$t$ at~$\Sigma_t$	(compatible with the boundary conditions at the intersection~$\partial \Sigma_t = \Sigma_t \cap \partial M$), there exists an unique solution to the Klein-Gordon equation in the whole spacetime satisfying these conditions.
		
	\item[C.]
		Using the temporal function as a coordinate, the metric should be written as
\begin{equation}
\rmd s^2 = - \rmd t^2 + h_{i j} (t, \vec{x}) \rmd x^i \rmd x^j,
\label{metric}
\end{equation}
where~$h_{i j} (t)$ is a regular Riemannian metric.\footnote{From here on we omit the explicit dependence of~$h_{i j} (t)$ and other quantities on the spatial coordinates. Also, in~(\ref{metric}) for simplicity we are assuming that for each hypersurface~$\Sigma_t$ there is one coordinate chart~$(x^1,\ldots,x^N)$ that completely covers it. This might not be the case, but considering several coordinate charts would be straightforward and would not affect the construction of the method.} This is called a \emph{synchronous gauge.}

\end{enumerate}

The necessity of each condition will become clear when constructing the method. A detailed discussion on their physical meaning and the limitations they introduce can be found in Appendix~A of Part~I. In this second article, we consider the cases in which at least some parts of the boundary~$\partial M$ have non-zero velocity in the coordinates chosen. We shall mention that, if the boundaries are not static, an alternative solution could be to change to a new coordinate system in which the boundaries remain static, and then use the method as exposed in Part~I. However, in general, in this new coordinate system the metric may not look like~(\ref{metric}) and (since Condition~C is also a requirement in Part~I) the integration method will not apply.

\subsection{Space of initial conditions at~$\Sigma_t$, inner product and self-adjoint operator}\label{objects}

Let us introduce three mathematical objects that are pivotal for the method. First, we define~$\Gamma_t$ as a subspace of the space of \emph{pairs of} square integrable smooth functions over a Cauchy hypersurface, representing possible initial conditions $(\Phi, \partial_t \Phi)|_{\Sigma_t}$. That is, $\Gamma_t \subset [C^{\infty} (\Sigma_t) \cap L^2 (\Sigma_t)]^{\oplus 2}$. Specifically, $\Gamma_t$ is the restriction of the full space of pairs of functions to initial conditions~$(\Phi, \partial_t \Phi)|_{\Sigma_t}$ satisfying Neumann vanishing boundary conditions~(\ref{neumann_cov}) at~$\partial \Sigma_t$. This can be rewritten in terms of the initial conditions as
\begin{equation}
\left\{
\begin{array}{l}
	\vec{n} \cdot \nabla_{h(t)} \Phi (t, \vec{x}) = - v_{\mathrm{B}} (t, \vec{x}) \partial_t \Phi (t, \vec{x}), \\
	\text{and}\ \vec{n} \cdot \nabla_{h(t)} \partial_t \Phi (t, \vec{x}) = 0\ \text{if}\ v_{\mathrm{B}}(t, \vec{x}) = 0;
\end{array}
\right.
\label{neumann_initial}
\end{equation}
where~$\vec{x} \in \partial \Sigma_t$; $\vec{n}(t, \vec{x})$ is the normal vector to~$\partial \Sigma_t$ and pointing outwards from~$\Sigma_t$; $v_{\mathrm{B}} (t, \vec{x}) := (\vec{n} \cdot \vec{v}_{\mathrm{B}}) (t, \vec{x})$, where $\vec{v}_{\mathrm{B}} (t, \vec{x})$ is the velocity vector of the boundary, and therefore~$v_{\mathrm{B}}(t, \vec{x})$ is its normal component; and~$\nabla_{h(t)}$ is the covariant derivative corresponding to the spatial metric~$h_{i j} (t)$. The first line of~(\ref{neumann_initial}) is just the reformulation of~(\ref{neumann_cov}) separating the spatial and temporal partial derivatives. The second line corresponds to the total time derivative of~(\ref{neumann_cov}) along the boundary when $v_{\mathrm{B}} = 0$. Clearly, that time derivative must also vanish, and this shall be taken into account when considering the compatibility of the initial conditions with the boundary conditions. However, in the regions where the boundary is moving in the chosen coordinates ($v_{\mathrm{B}} \neq 0$), the total time derivative of~(\ref{neumann_cov}) along the boundary involves second order partial time derivatives of the field. In such case, the fulfilment of the first time derivative of the boundary conditions at~$\partial \Sigma_t$ already depends on the dynamical evolution given by the Klein-Gordon equation~(\ref{klein-gordon}), and not just on the initial conditions at~$\Sigma_t$. Therefore, its fulfilment is guaranteed by Condition~B. On the other hand, in the regions where the boundary remains parallel to~$\partial_t$ ($v_{\mathrm{B}} = 0$), the first time derivative of the boundary condition reads as in the second line of~(\ref{neumann_initial}), involving only up to the first partial time derivative of the field, and thus depending only on the initial conditions at~$\Sigma_t$. Therefore, in such regions it imposes the corresponding constraint on these initial conditions.

The second mathematical object that we introduce is the following inner product in the Hilbert space of pairs of functions $L^2(\Sigma_t) \oplus L^2(\Sigma_t) \supset \Gamma_t$:
\begin{multline}
\left\langle
\left( \begin{array}{c}
	\Phi' \\
	\partial_t \Phi'
\end{array} \right)
,
\left( \begin{array}{c}
	\Phi \\
	\partial_t \Phi
\end{array} \right)
\right\rangle_{\Sigma_t}
:= \\
\int_{\Sigma_t} \rmd V_t\ [\xi R^h (t) + m^2 + F(t)] \Phi' \Phi^*\\
+ \int_{\Sigma_t} \rmd V_t\ \partial_t \Phi' \partial_t \Phi^* + \int_{\Sigma_t} \rmd V_t\ (\nabla_{h(t)} \Phi') \cdot (\nabla_{h(t)} \Phi^*);
\label{scalar_product_ini}
\end{multline}
where~$R^{h}(t)$ is the scalar curvature corresponding to the spatial metric~$h_{i j} (t)$ and $F(t) \geq 0$ a time-dependent non-negative quantity given by
\begin{multline}
F(t) := \\
\left\{
\begin{array}{l}
	0 \quad \text{if}\ \xi R^h (t, \vec{x}) + m^2 > 0\ \text{a.e.\ in}\ \Sigma_t,\\
	-\text{ess\ inf} \{\xi R^h (t, \vec{x}) + m^2, \vec{x} \in \Sigma_t\} + \epsilon \quad \text{i.o.c.};
\end{array}
\right.
\label{def_F}
\end{multline}
where ``a.e.'' stands for \emph{almost everywhere,} ``ess inf'' stands for \emph{essential infimum,} ``i.o.c.'' stands for \emph{in other case} and $\epsilon > 0$ is an arbitrarily small positive quantity. The quantity~$F(t)$ ensures that the inner product is positive-definite.\footnote{The quantity~$F(t)$ plays here an analogous role to that of the function with the same name introduced in Part~I, which meaning is discussed in Appendix~B there. In many problems of interest, such as a massive field with minimal coupling ($\xi = 0$), the quantity~$F(t)$ vanishes and can be ignored.}

Finally, we introduce the following operator in~$\Gamma_t$:
\begin{equation}
\hat{\mathscr{M}}(t) :=
\left( \begin{array}{cc}
0	& ~1~ \\
	\hat{\mathscr{O}}(t) & ~0~
\end{array} \right);
\label{operator_M}
\end{equation}
with
\begin{equation}
\hat{\mathscr{O}}(t) := - \nabla_{h(t)}^2 + \xi R^h(t) + m^2 + F(t),
\label{operator}
\end{equation}
where~$\nabla_{h(t)}^2$ is the Laplace-Beltrami differential operator and~$F(t)$ has been defined in~(\ref{def_F}). With the inner product in~(\ref{scalar_product_ini}), the boundary condition~(\ref{neumann_initial}), and using Green's first identity, one can easily check that the operator~$\hat{\mathscr{M}}(t)$ is self-adjoint.\footnote{Assertions about the self-adjoint nature of the operator should strictly be done over the extension of the operator defined on~$\Gamma_t$ to the full Hilbert space~$L^2 (\Sigma_t) \oplus L^2(\Sigma_t)$ (of which~$\Gamma_t$ is a dense subspace). This is a well-known mathematical procedure in the analysis of partial differential equations with elliptic operators and boundary value problems. Since this is the context in which we make use of the operator, we shall not get into the details of it. See for example~\cite{taylor1996partial}.}

Before finishing this Section, let us mention that using the operator~$\hat{\mathscr{O}}(t)$ and Condition~C we shall simplify the Klein-Gordon equation~(\ref{klein-gordon}) taking into account the form of the metric in~(\ref{metric}), obtaining
\begin{equation}
\partial_t^2 \Phi = - \hat{\mathscr{O}}(t) \Phi - q(t) \partial_t \Phi - \xi \bar{R}(t) \Phi + F(t) \Phi;
\label{klein-gordon_2}
\end{equation}
where
\begin{equation}
q(t) := \partial_t \log \sqrt{h(t)}
\label{change_factor}
\end{equation}
is a factor which depends on the change of the metric of the spacelike hypersurfaces with time, with~$h(t)$ the determinant of the spatial metric~$h_{i j} (t)$, and $\bar{R}(t) := R(t) - R^{h}(t)$ is the part of the full scalar curvature of~$g_{\mu \nu}$ which depends on time derivatives, given by
\begin{equation}
\bar{R}(t) = 2 \partial_t q(t) + q(t)^2 - \frac{1}{4} [\partial_t h^{i j} (t)] [\partial_t h_{i j} (t)].
\label{bar_scalar}
\end{equation}
The key role of Condition~C has been to yield equation~(\ref{klein-gordon_2}), in which all the spatial derivatives present are those in the Laplace-Beltrami operator contained in~$\hat{\mathscr{O}}(t)$.

\section{Construction of the method}\label{nuclear}

\subsection{Construction of the bases of modes}\label{bases}

For each spacelike hypersurface $\Sigma_{\tilde{t}}$ we construct a set of modes~$\{\prescript{\pm}{}\Phi^{[\tilde{t}]}_n(t)\}$ fulfilling the following two Properties:

\begin{enumerate}
	\item[I.]
		They form a complete basis of the space of solutions to the Klein-Gordon equation~(\ref{klein-gordon}). We stress that each mode of the basis is defined in the whole spacetime, the label~$[\tilde{t}]$ meaning only that we \emph{associate} it to the corresponding hypersurface. Specifically, it is in this hypersurface that we set its initial conditions.
		
	\item[II.]
		If there exists a region~$S$ of the spacetime around~$\Sigma_{\tilde{t}}$ where
		\begin{itemize}
		\item $\partial_t$ behaves like a Killing field [$h_{i j} (t)$ is constant],
		
		\item the boundaries remain parallel to~$\partial_t$ (``static''),

		\item and the function~$F(t)$ in~(\ref{def_F}) vanishes;
		\end{itemize}
		the modes form an orthonormal basis with respect to the Klein-Gordon inner product~(\ref{scalar_product}) of positive frequency modes (modes~$\prescript{+}{}\Phi^{[\tilde{t}]}_n$) and negative frequency modes (modes~$\prescript{-}{}\Phi^{[\tilde{t}]}_n$) with respect to~$t$.\footnote{Regions in which~$F(t)$ does not vanish, but takes an arbitrarily small value~$\epsilon$, may also be considered. We refer to Appendix~B of Part~I for further details on this condition.} The region~$S$ needs to fully embrace the spatial hypersurfaces~$\Sigma_t$ along an interval of time~$t$ around~$t=\tilde{t}$ which is long enough so as to explore the minimum frequency in the spectrum given by the modes.
		
\end{enumerate}

\begin{figure}[h]
\begin{center}
\includegraphics[width=6cm]{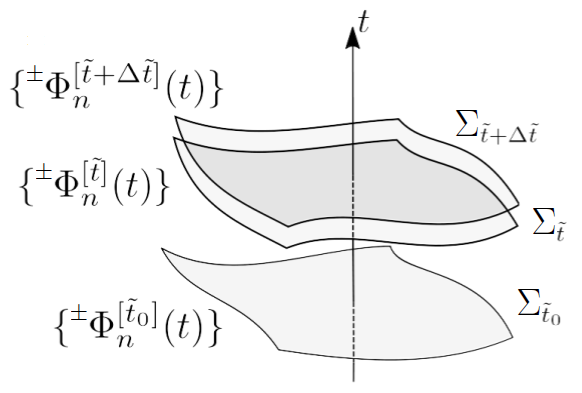}
\caption{Association of bases of modes~$\{\prescript{\pm}{}\Phi^{[\tilde{t}]}_n (t)\}$ to Cauchy hypersurfaces~$\Sigma_{\tilde{t}}$.}
\label{slices}
\end{center}
\end{figure}

In Fig.~\ref{slices} we provide a graphical depiction of this association of bases of modes to Cauchy hypersurfaces, which shall be helpful when following the construction of the method. In order to construct the bases $\{\prescript{\pm}{}\Phi^{[\tilde{t}]}_n\}$, we first construct auxiliary bases of modes $\{\Psi^{[\tilde{t}]}_n\}$, also associated to each hypersurface~$\Sigma_{\tilde{t}}$, and then introduce a linear transformation to the bases $\{\prescript{\pm}{}\Phi^{[\tilde{t}]}_n\}$. Since the modes~$\Psi^{[\tilde{t}]}_n (t)$ are also solutions to the Klein-Gordon equation, because of Condition~B the only quantities left to fully determine them are the initial conditions, which we are going to fix at~$\Sigma_{\tilde{t}}$. That is, we need to fix the pair $(\Psi^{[\tilde{t}]}_n (\tilde{t}), \partial_t \Psi^{[\tilde{t}]}_n (t)|_{t=\tilde{t}}) \in \Gamma_{\tilde{t}}$ for each mode. These pairs are going to be given by the eigenvectors of the operator~$\hat{\mathscr{M}}(\tilde{t})$ in~(\ref{operator_M}):
\begin{equation}
\hat{\mathscr{M}}(\tilde{t})
\left( \begin{array}{c}
	\Psi^{[\tilde{t}]}_n (\tilde{t}) \\
	\partial_t \Psi^{[\tilde{t}]}_n (t) |_{t=\tilde{t}}
\end{array} \right)
=
\omega^{[\tilde{t}]}_n
\left( \begin{array}{c}
	\Psi^{[\tilde{t}]}_n (\tilde{t}) \\
	\partial_t \Psi^{[\tilde{t}]}_n (t) |_{t=\tilde{t}}
\end{array} \right).
\label{eigenvalues_good}
\end{equation}
Notice that $\tilde{t}$ in~(\ref{eigenvalues_good}) is just a parameter.\footnote{As we commented previously for the operator~$\hat{\mathscr{M}}(\tilde{t})$, this equation should also be posed in the full Hilbert space~$L^2 (\Sigma_t) \oplus L^2 (\Sigma_t)$. It is known, however, that the functions representing the eigenvectors can be taken to belong to the dense subspace~$\Gamma_{\tilde{t}}$ \cite{taylor1996partial}.} Since~$\hat{\mathscr{M}}(\tilde{t})$ is self-adjoint with respect to the inner product~(\ref{scalar_product_ini}), the eigenvalues~$\omega^{[\tilde{t}]}_n$ are real. Therefore, we can also impose that the pairs of functions are of real functions and that they satisfy the following orthogonality and normalisation condition:
\begin{equation}
\left\langle
\left( \begin{array}{c}
	\Psi^{[\tilde{t}]}_n (\tilde{t}) \\
	\partial_t \Psi^{[\tilde{t}]}_n (t) |_{t=\tilde{t}}
\end{array} \right)
,
\left( \begin{array}{c}
	\Psi^{[\tilde{t}]}_m (\tilde{t}) \\
	\partial_t \Psi^{[\tilde{t}]}_m (t) |_{t=\tilde{t}}
\end{array} \right)
\right\rangle_{\Sigma_{\tilde{t}}}
= |\omega^{[\tilde{t}]}_n| \delta_{n m}.
\label{norma_casera}
\end{equation}
In \ref{no_zeros} we prove that there are no zero eigenvalues~$\omega^{[\tilde{t}]}_n$, so the normalisation criterion is valid. Finally, notice that thanks to Condition~A we can be sure that the spectrum is discrete, as we had implicitly assumed with the notation.

In order to operationally find the quantities~$\Psi^{[\tilde{t}]}_n (\tilde{t})$, we summarise~(\ref{neumann_initial}) and~(\ref{eigenvalues_good}) in the two equations
\begin{align}
\hat{\mathscr{O}}(\tilde{t}) \Psi^{[\tilde{t}]}_n (\tilde{t}) & = (\omega^{[\tilde{t}]}_n)^2 \Psi^{[\tilde{t}]}_n (\tilde{t}), \label{neumann_moving_modes} \\
\vec{n} \cdot \nabla_{h(\tilde{t})} \Psi^{[\tilde{t}]}_n(\tilde{t}, \vec{x}) & = - \omega^{[\tilde{t}]}_n v_{\mathrm{B}} (\tilde{t}, \vec{x}) \Psi^{[\tilde{t}]}_n (\tilde{t}, \vec{x}), \quad \vec{x} \in \partial \Sigma_{\tilde{t}};
\label{neumann_moving_modes_boundary}
\end{align}
and the equation for the partial time derivative
\begin{equation}
\partial_t \Psi^{[\tilde{t}]}_n(t) |_{t=\tilde{t}} = \omega^{[\tilde{t}]}_n \Psi^{[\tilde{t}]}_n(\tilde{t}).
\label{first_derivative}
\end{equation}
This last equation plays a role in the construction of the method, but in order to apply the method to a concrete problem only the first two are necessary. Because of this last equation, when~$v_{\mathrm{B}}(\tilde{t}, \vec{x}) = 0$ equation~(\ref{neumann_moving_modes_boundary}) is also imposing the second condition in~(\ref{neumann_initial}). Notice that equation~(\ref{neumann_moving_modes}) cannot be taken as an eigenvalue problem posed directly for~$\Psi^{[\tilde{t}]}_n (\tilde{t})$, since the boundary conditions~(\ref{neumann_moving_modes_boundary}) of such problem would not be fixed (they would depend on the eigenvalue).

By Condition~B, for each pair $(\Psi^{[\tilde{t}]}_n (\tilde{t}), \partial_t \Psi^{[\tilde{t}]}_n (t)|_{t=\tilde{t}}) \in \Gamma_{\tilde{t}}$ that is solution to~(\ref{eigenvalues_good}) we have an unique mode~$\Psi^{[\tilde{t}]}_n (t)$. Let us now relabel the modes in the (infinite countable) set~$\{\Psi^{[\tilde{t}]}_n\}$ and group them into two (also infinite countable) subsets, $\{\prescript{+}{}\Psi^{[\tilde{t}]}_n\}$ and~$\{\prescript{-}{}\Psi^{[\tilde{t}]}_n\}$, with their respective sets of eigenvalues~$\{\prescript{+}{}\omega^{[\tilde{t}]}_n\}$ and~$\{\prescript{-}{}\omega^{[\tilde{t}]}_n\}$. In \ref{props_bases} we prove that, at least for the problems in which we can give a physical interpretation to the results of the method, there is an infinite number of both positive and negative eigenvalues~$\omega^{[\tilde{t}]}_n$. We organise the solutions as follows:
\begin{equation}
\cdots \leq \prescript{-}{}\omega^{[\tilde{t}]}_2 \leq \prescript{-}{}\omega^{[\tilde{t}]}_1 < 0 < \prescript{+}{}\omega^{[\tilde{t}]}_1 \leq \prescript{+}{}\omega^{[\tilde{t}]}_2 \leq \cdots.
\label{eig_org}
\end{equation}
We notice that, in general, the solutions~$\prescript{\pm}{}\Psi^{[\tilde{t}]}_n$ and eigenvalues~$\prescript{\pm}{}\omega^{[\tilde{t}]}_n$ in each subset are independent.

Because of Condition~B and the linearity of the Klein-Gordon equation, we can trivially consider the inner product~(\ref{scalar_product_ini}), defined for initial conditions at each hypersurface $\Sigma_{\tilde{t}}$, as an inner product in the space of solutions: For any two solutions, their inner product is that of their corresponding initial conditions at the given hypersurface. Using this definition of inner product in the space of solutions, and carefully taking into account the relabelling $\Psi^{[\tilde{t}]}_n \to \prescript{\pm}{}\Psi^{[\tilde{t}]}_n$, we can write the orthonormalisation condition~(\ref{norma_casera}) in an equivalent way, but directly for the modes~$\prescript{\pm}{}\Psi^{[\tilde{t}]}_n$, as\footnote{Any time that two `$\pm$' signs are involved in an equation, we use a \emph{hat} `$\hat{\pm}$' to distinguish one of them.}
\begin{equation}
\langle \prescript{\pm}{}\Psi^{[\tilde{t}]}_n, \prescript{\hat{\pm}}{}\Psi^{[\tilde{t}]}_m \rangle_{\Sigma_{\tilde{t}}} = |\prescript{\pm}{}\omega^{[\tilde{t}]}_n| \delta_{n m} \delta_{\pm \hat{\pm}},
\label{norma_casera_psi}
\end{equation}
where the quantity~$\delta_{\pm \hat{\pm}}$ equals~$1$ if the signs coincide and~$0$ otherwise. We also stress that this orthonormalisation is only correct in the inner product of the hypersurface~$\Sigma_{\tilde{t}}$ to which the modes are associated.

Finally, we build the set of modes~$\{\prescript{\pm}{}\Phi^{[\tilde{t}]}_n\}$ by taking a linear transformation from the set of modes~$\{\prescript{\pm}{}\Psi^{[\tilde{t}]}_n\}$. This linear transformation reads
\begin{equation}
\left( \begin{array}{c}
	\prescript{+}{}\Phi^{[\tilde{t}]}_1 \\
	\ldots \\
	\prescript{-}{}\Phi^{[\tilde{t}]}_1 \\
	\ldots
\end{array} \right)
= M
\left( \begin{array}{c}
	\prescript{+}{}\Psi^{[\tilde{t}]}_1 \\
	\ldots \\
	\prescript{-}{}\Psi^{[\tilde{t}]}_1 \\
	\ldots
\end{array} \right),
\label{psi_to_phi}
\end{equation}
where, in obvious block notation,
\begin{equation}
M := \frac{1}{2}
\left( \begin{array}{cc}
	(1-\rmi) I & (1+\rmi) I \\
	(1+\rmi) I & (1-\rmi) I
\end{array} \right).
\label{matrix_M}
\end{equation}

The set of modes~$\{\prescript{\pm}{}\Phi^{[\tilde{t}]}_n\}$, constructed this way for each~$\tilde{t}$, is a basis of modes satisfying Properties~I and~II. We prove this in \ref{props_bases}. In particular, in the regions~$S$ described in Property~II, where it is possible to construct modes with well-defined frequency with respect to~$t$, we have that $\prescript{-}{}\Psi^{[\tilde{t}]}_n (\tilde{t}) = \prescript{+}{}\Psi^{[\tilde{t}]}_n (\tilde{t})$ and $\prescript{-}{}\omega^{[\tilde{t}]}_n = - (\prescript{+}{}\omega^{[\tilde{t}]}_n)$, and the modes in the basis are
\begin{equation}
\prescript{\pm}{}\Phi^{[\tilde{t}]}_n (t) = \prescript{+}{}\Psi^{[\tilde{t}]}_n (\tilde{t}) \rme^{\mp \rmi (\prescript{+}{}\omega^{[\tilde{t}]}_n) (t-\tilde{t})};
\label{modes_with_freq}
\end{equation}
that is, $\prescript{+}{}\Phi^{[\tilde{t}]}_n (t)$ are the modes with positive frequencies $\prescript{+}{}\omega^{[\tilde{t}]}_n > 0$, and $\prescript{-}{}\Phi^{[\tilde{t}]}_n (t) = \prescript{+}{}\Phi^{[\tilde{t}]}_n (t)^*$ the corresponding negative frequency modes.

The construction of the bases of modes done here has been significantly different to that in Part~I. We discuss in~\ref{why_part_II} why the construction done in Part~I does not work here.

\subsection{Time-dependent linear transformation}\label{sec_linear_transf}

Let us write down formally the linear transformation $U (\tilde{t}, \tilde{t}_0)$ between any two bases of modes, associated to the hypersurfaces~$\Sigma_{\tilde{t}_0}$ and~$\Sigma_{\tilde{t}}$:
\begin{equation}
\left( \begin{array}{c}
	\prescript{+}{}\Phi^{[\tilde{t}]}_1 \\
	\ldots \\
	\prescript{-}{}\Phi^{[\tilde{t}]}_1 \\
	\ldots
\end{array} \right)
= U (\tilde{t}, \tilde{t}_0)
\left( \begin{array}{c}
	\prescript{+}{}\Phi^{[\tilde{t}_0]}_1 \\
	\ldots \\
	\prescript{-}{}\Phi^{[\tilde{t}_0]}_1 \\
	\ldots
\end{array} \right).
\label{transformation_moving_U}
\end{equation}
In \ref{great_proof} we prove that this time-dependent linear transformation between bases satisfies the differential equation
\begin{equation}
\frac{\rmd}{\rmd \tilde{t}} U(\tilde{t}, \tilde{t}_0) = M \hat{V}(\tilde{t}) M^* U(\tilde{t}, \tilde{t}_0);
\label{differential_equation_moving_U}
\end{equation}
where
\begin{equation}
\hat{V}(\tilde{t}) =
\left(
\begin{array}{cc}
	\hat{V}^{++} & \hat{V}^{+-} \\
	\hat{V}^{-+} & \hat{V}^{--}
\end{array}
\right),
\label{def_Vhat}
\end{equation}
with
\begin{multline}
\hat{V}^{\pm \hat{\pm}}_{nm} = -(\prescript{\hat{\pm}}{}\omega_m^{[\tilde{t}]})\delta_{nm}\delta_{\pm \hat{\pm}} \\
\hat{\pm} \left\{ \left[(\prescript{\pm}{}\omega_n^{[\tilde{t}]})+(\prescript{\hat{\pm}}{}\omega_m^{[\tilde{t}]})\right]\int_{\Sigma_{\tilde{t}}}\rmd V_{\tilde{t}} \left[\frac{\rmd}{\rmd \tilde{t}}\prescript{\pm}{}\Psi_n^{[\tilde{t}]}(\tilde{t}) \right] \prescript{\hat{\pm}}{}\Psi_m^{[\tilde{t}]}(\tilde{t}) \right. \\
\left.+\left[2 (\prescript{\pm}{}\omega_n^{[\tilde{t}]} )^2+\frac{\rmd\prescript{\pm}{}\omega_n^{[\tilde{t}]}}{\rmd \tilde{t}} - F(\tilde{t})\right]\int_{\Sigma_{\tilde{t}}}\rmd V_{\tilde{t}}\prescript{\pm}{}\Psi_n^{[\tilde{t}]}(\tilde{t})\prescript{\hat{\pm}}{}\Psi_m^{[\tilde{t}]}(\tilde{t}) \right. \\
\left.+\int_{\Sigma_{\tilde{t}}}\rmd V_{\tilde{t}}\prescript{\pm}{}\Psi_n^{[\tilde{t}]}(\tilde{t})\left[\prescript{\pm}{}\omega_n^{[\tilde{t}]}q(\tilde{t})+\xi\bar{R}(\tilde{t})\right]\prescript{\hat{\pm}}{}\Psi_m^{[\tilde{t}]}(\tilde{t}) \right. \\
\left. - \int_{\partial \Sigma_{\tilde{t}}}\rmd S_{\tilde{t}}\ v_{\mathrm{B}}(\tilde{t}) \left[\frac{\rmd}{\rmd \tilde{t}}\prescript{\pm}{}\Psi_n^{[\tilde{t}]}(\tilde{t}) \right] \prescript{\hat{\pm}}{}\Psi_m^{[\tilde{t}]}(\tilde{t}) \right\},
\label{V_hat}
\end{multline}
where~$\rmd S_{\tilde{t}}$ is the surface element of~$\partial \Sigma_{\tilde{t}}$. With the initial condition~$U(\tilde{t}_0, \tilde{t}_0) = I$, equation (\ref{differential_equation_moving_U}) has the formal solution
\begin{equation}
U(\tilde{t}_{\mathrm{f}}, \tilde{t}_0) = \mathscr{T} \exp \left[ \int_{\tilde{t}_0}^{\tilde{t}_{\mathrm{f}}} \rmd \tilde{t}\ M \hat{V}(\tilde{t}) M^* \right],
\label{U_moving_solution}
\end{equation}
where $\mathscr{T}$ denotes time ordering.

Equations~(\ref{differential_equation_moving_U}-\ref{U_moving_solution}) are one of the two main results of this work: They switch from the time evolution of the modes to the time evolution of the transformation between bases. The time-dependent linear transformation obtained relates the bases~$\{\prescript{\pm}{}\Phi^{[\tilde{t}]}_n\}$, which are those satisfying Properties~I and~II. However, one of the strengths of the method is that all the quantities appearing in~(\ref{V_hat}), which are the coefficients of our differential equation, are known just by computing the initial conditions of the auxiliary bases~$\{\prescript{\pm}{}\Psi^{[\tilde{t}]}_n\}$, which are the solutions to the equations~(\ref{neumann_moving_modes}) and~(\ref{neumann_moving_modes_boundary}), for which the time~$\tilde{t}$ is just a parameter.

\subsection{Physical interpretation}\label{physical}

The time-dependent linear transformation obtained contains all the information necessary to compute the evolution of the field in time. However, as we advanced in the Introduction, we do not pretend to give a quantisation for each and every basis of modes that we have constructed at each time. It is only in those regions~$S$ described within Property~II of Subsect.~\ref{bases}, where we can proceed to the usual Fock quantisation of the field. That is, defining the corresponding Fock space with its vacuum state and creation and annihilation operators (ones the adjoints of the others in the case of a real field) associated to the mode decomposition given by the method. This is the case because, in such regions, the decomposition is done in modes with well-defined frequency with respect to a timelike Killing field [see equation~(\ref{modes_with_freq})]. We rely on the fact that, in such situation, Fock representation gives the correct physical description of a field in terms of particles associated to those modes.\footnote{For brevity, we do not expose the details of the quantisation procedure explicitly, see e.g.~\cite{BD1984}.} When connecting two regions where this Fock quantisation procedure can be done, the time-dependent linear transformation that we constructed in the previous Subsection really becomes a Bogoliubov transformation, taking the well-known form
\begin{equation}
U(\tilde{t}, \tilde{t}_0) =
\left(
\begin{array}{cc}
	\alpha (\tilde{t}, \tilde{t}_0) & \beta (\tilde{t}, \tilde{t}_0) \\
	\beta (\tilde{t}, \tilde{t}_0)^* & \alpha (\tilde{t}, \tilde{t}_0)^*
\end{array}
\right).
\label{bogoliubov_matrix}
\end{equation}
These Bogoliubov coefficients also relate in the usual way the annihilation and creation operators of the mode decompositions associated to the different regions.

We refer to Subsect.~3.3 of Part~I for additional discussions on different aspects of the physical interpretation, which also apply here.

\section{Small perturbations and resonances}\label{res}

Let us consider the case in which the spatial metric~$h_{ij} (t)$ only changes in time by a small perturbation around some constant metric~$h^0_{ij}$; that is,
\begin{equation}
h_{ij} (t) = h^0_{ij} + \varepsilon \Delta h_{ij} (t),
\label{perturbation_metric}
\end{equation}
where~$\varepsilon \ll 1$. Also, the boundaries may experience small displacements of order~$\varepsilon$, meaning that we allow the hypersurfaces~$\Sigma_t$ to slightly change around some fixed hypersurface~$\Sigma^0$. We call~$\varepsilon \Delta x (t, \vec{x})$ the proper distance between the boundary~$\partial \Sigma_t$ and the fixed boundary~$\partial \Sigma^0$ at the point~$(t,\vec{x})$ along the direction normal and outwards to~$\partial \Sigma^0$. Therefore, we have that
\begin{equation}
v_{\mathrm{B}} (t, \vec{x}) \approx \varepsilon \frac{\rmd}{\rmd t} \Delta x(t, \vec{x}).
\label{vB_x}
\end{equation}
Finally, we require that~$F(t)$ remains~$O(\varepsilon)$, so that the solutions to the problem for~$\varepsilon = 0$ are modes with well-defined frequency. In \ref{simplif} we prove that such value of~$F(t)$ actually does not contribute at all to the physically relevant result of resonances. Therefore, once we have required that~$F(t)$ remains~$O(\varepsilon)$, without loss of generality we consider that $F(t)=0$.

We write down the quantities in~(\ref{operator}) and~(\ref{bar_scalar}) to first order in~$\varepsilon$:
\begin{align}
\hat{\mathscr{O}} (\tilde{t}) & \approx \hat{\mathscr{O}}^0 + \varepsilon \Delta\hat{\mathscr{O}} (\tilde{t}), \nonumber \\
\bar{R}(\tilde{t}) & \approx \varepsilon \Delta \bar{R}(\tilde{t}); \label{perturbation_quantities}
\end{align}
where $\bar{R}(\tilde{t})$ vanishes when there is no perturbation because it only depends on time derivatives. The perturbation of the quantity~$q(\tilde{t})$ in~(\ref{change_factor}) does not directly appear in the perturbative regime, but rather its primitive with respect to~$\tilde{t}$, given by
\begin{equation}
\Delta r (\tilde{t}) := \frac{1}{2} \left. \frac{\partial}{\partial \varepsilon} \log h(\tilde{t}) \right|_{\varepsilon = 0}.
\label{def_dr}
\end{equation}

As we prove in \ref{simplif}, we manage to write down the Bogoliubov coefficients without explicitly computing the solutions~$\prescript{\pm}{}\Psi^{[\tilde{t}]}_n (\tilde{t})$ and~$\prescript{\pm}{}\omega_n^{[\tilde{t}]}$ to first order in~$\varepsilon$, by using the perturbation of the operator~$\Delta\hat{\mathscr{O}} (\tilde{t})$ in~(\ref{perturbation_quantities}). Thus, we only need the solutions to~(\ref{neumann_moving_modes}) and~(\ref{neumann_moving_modes_boundary}) for the static problem (for $\varepsilon = 0$). We denote them as~$\Psi^0_n$ and~$\omega^0_n > 0$ for the~$\prescript{+}{}\Psi^{[\tilde{t}]}_n$ modes. Therefore, those corresponding to the~$\prescript{-}{}\Psi^{[\tilde{t}]}_n$ modes are~$\Psi^0_n$ and~$-\omega^0_n$ (see \ref{props_bases}). These solutions satisfy
\begin{align}
\hat{\mathscr{O}}^0 \Psi^0_n & = (\omega^0_n)^2 \Psi^0_n,
\label{op_zeroth} \\
\vec{n} \cdot \nabla_{h^0} \Psi^0_n(\vec{x}) & = 0, \quad \vec{x} \in \partial \Sigma^0.
\label{boundary_zeroth}
\end{align}
Being solutions to a static problem, they must also fulfil the orthonormalisation condition given in~(\ref{psi_norm}), namely,
\begin{equation}
\int_{\Sigma^0} \rmd V^0\ \Psi^0_n \Psi^0_m = \frac{\delta_{n m}}{2 \omega^0_n},
\label{psi_zero_norm}
\end{equation}
where~$\rmd V^0$ is the volume element of~$\Sigma^0$.

We want to solve the differential equation~(\ref{differential_equation_moving_U}) in the perturbative regime. Let us first compute the coefficient $M \hat{V} (\tilde{t}) M^*$ explicitly to zeroth order in~$\varepsilon$. Using the solutions to zeroth order and~(\ref{psi_zero_norm}) in~(\ref{V_hat}) it is easy to check that
\begin{equation}
\hat{V}^{\pm \hat{\pm}}_{nm} = \hat{\pm} \omega_n^0 \delta_{n m} \delta_{\mp \hat{\pm}} + O (\varepsilon).
\label{V_hat_zeroth}
\end{equation}
Using now~(\ref{matrix_M}) and~(\ref{def_Vhat}), we can write
\begin{equation}
M \hat{V} (\tilde{t}) M^* \approx \rmi \Omega^0 + \varepsilon \Delta K (\tilde{t});
\label{lambda_zeroth}
\end{equation}
with
\begin{align}
\Omega^0 &:= \diag (\omega^0_1, \omega^0_2, \ldots, -\omega^0_1, -\omega^0_2, \ldots),\nonumber \\
\Delta K(\tilde{t}) &:=
\left(
\begin{array}{cc}
	\Delta \hat{\alpha} (\tilde{t}) & \Delta \hat{\beta} (\tilde{t}) \\
	\Delta \hat{\beta} (\tilde{t})^* & \Delta \hat{\alpha} (\tilde{t})^*
\end{array}
\right),
\label{perturbation_matrix}
\end{align}
where the entries of~$\Delta K(\tilde{t})$ are the contributions to first order in~$\varepsilon$, that depend on the perturbation and which explicit expressions we provide later on. If we introduce the result~(\ref{lambda_zeroth}) in the differential equation~(\ref{differential_equation_moving_U}), we clearly see that for $\varepsilon = 0$ the modes evolve just with a trivial phase with constant frequency $\pm \omega^0_n$, as one should expect for a static metric.

In order to properly compute the evolution to first order in~$\varepsilon$, we shall first absorb any phase evolution, given by the diagonal terms. This is done by writing the evolution in terms of a new linear transformation~$Q(\tilde{t}, \tilde{t}_0)$ defined by
\begin{equation}
Q(\tilde{t}, \tilde{t}_0) := \Theta(\tilde{t})^* U(\tilde{t}, \tilde{t}_0);
\label{Q_definition}
\end{equation}
where
\begin{align}
\Theta(\tilde{t}) & := \exp \left\{ \int^{\tilde{t}} \rmd \tilde{t}' [\rmi \Omega^0 + \varepsilon \Delta A(\tilde{t}')] \right\}, \label{exponential_omega} \\
\Delta A(\tilde{t}) & := \diag(\Delta \hat{\alpha}_{1 1}, \Delta \hat{\alpha}_{2 2}, \ldots, -\Delta \hat{\alpha}_{1 1}, -\Delta \hat{\alpha}_{2 2}, \ldots). \nonumber
\end{align}
Replacing~(\ref{Q_definition}) in~(\ref{differential_equation_moving_U}), we get the differential equation
\begin{align}
\frac{\rmd}{\rmd \tilde{t}} Q(\tilde{t}, \tilde{t}_0) = &\ \varepsilon \Theta^0(\tilde{t})^* \Delta \bar{K}(\tilde{t}) \Theta^0(\tilde{t}) Q(\tilde{t}, \tilde{t}_0), \label{Q_differential_equation} \\
\Delta \bar{K}(\tilde{t}) := &\ \Delta K(\tilde{t}) - \Delta A(\tilde{t}), \nonumber\\
\Theta^0(\tilde{t}) := &\ \rme^{\rmi \Omega^0 \tilde{t}};
\nonumber
\end{align}
where we dropped the terms to first order in~$\varepsilon$ from~$\Theta(\tilde{t})$ because of the overall factor~$\varepsilon$ appearing on the r.h.s. With the initial condition $Q(\tilde{t}_0, \tilde{t}_0) = I$, to first order in~$\varepsilon$ the transformation reads
\begin{equation}
Q(\tilde{t}_{\mathrm{f}}, \tilde{t}_0) \approx I + \varepsilon \int_{\tilde{t}_0}^{\tilde{t}_{\mathrm{f}}} \rmd \tilde{t}\ \Theta^0(\tilde{t})^* \Delta \bar{K}(\tilde{t}) \Theta^0(\tilde{t}).
\label{perturbation_Q}
\end{equation}

We can show the resonance behaviour of the field in a clear way if we write explicitly the expressions for the Bogoliubov coefficients:
\begin{align}
\alpha_{n n} (\tilde{t}_{\mathrm{f}}, \tilde{t}_0) \approx &\ 1; \nonumber \\
\alpha_{n m} (\tilde{t}_{\mathrm{f}}, \tilde{t}_0) \approx &\ \varepsilon \int_{\tilde{t}_0}^{\tilde{t}_{\mathrm{f}}} \rmd \tilde{t}\ \rme^{-\rmi (\omega^0_n - \omega^0_m) \tilde{t}} \Delta \hat{\alpha}_{n m} (\tilde{t}), \label{perturbation_alpha} \\
& \ n \neq m; \nonumber \\
\beta_{n m} (\tilde{t}_{\mathrm{f}}, \tilde{t}_0) \approx &\ \varepsilon \int_{\tilde{t}_0}^{\tilde{t}_{\mathrm{f}}} \rmd \tilde{t}\ \rme^{-\rmi (\omega^0_n + \omega^0_m) \tilde{t}} \Delta \hat{\beta}_{n m} (\tilde{t}). \label{perturbation_beta}
\end{align}

In general, the Bogoliubov transformation differs from the identity just by terms of first order in~$\varepsilon$, except for the cases where there are resonances. That is, if the perturbation considered contains some characteristic frequency~$\omega_{\mathrm{p}}$, then the same frequency is usually also present in the quantities~$\Delta \hat{\alpha}_{n m} (\tilde{t})$ and~$\Delta \hat{\beta}_{n m} (\tilde{t})$. If such frequency coincides with some difference between the frequencies of two modes, $\omega_{\mathrm{p}} = \omega^0_n - \omega^0_m$ (it is in resonance), then the corresponding coefficient~$\alpha_{n m} (\tilde{t}_{\mathrm{f}}, \tilde{t}_0)$ grows linearly with the time difference~$\tilde{t}_{\mathrm{f}} - \tilde{t}_0$, and after enough time it will overcome the~$O(\varepsilon)$. Respectively, if the characteristic frequency coincides with some sum between the frequencies of two modes, $\omega_{\mathrm{p}} = \omega^0_n + \omega^0_m$, then the corresponding coefficient~$\beta_{n m} (\tilde{t}_{\mathrm{f}}, \tilde{t}_0)$ grows linearly in time and eventually overcomes the~$O(\varepsilon)$.\footnote{In Appendix~E.1 of Part~I we show that these resonances remain stable under small deviations of the frequency of the perturbation from the exact resonant frequency.}

If the Fourier transform~$\mathscr{F}$ of~$\Delta \hat{\alpha}_{n m} (\tilde{t})$ [respectively $\Delta \hat{\beta}_{n m} (\tilde{t})$] exists as a well-defined function, which necessarily implies that the perturbation vanishes fast enough in the asymptotic past and future, then another way to consider the resonances is by taking the limits~$\tilde{t}_0 \to -\infty$ and~$\tilde{t}_{\mathrm{f}} \to \infty$ in~(\ref{perturbation_alpha}) and~(\ref{perturbation_beta}) and writing
\begin{align}
\alpha_{n n} (-\infty,\infty) \approx &\ 1; \nonumber \\
\alpha_{n m} (-\infty,\infty) \approx &\ \varepsilon \sqrt{2 \pi}\ \mathscr{F} [\Delta \hat{\alpha}_{n m}] (\omega^0_n - \omega^0_m), \label{fourier_alpha} \\
& \ n \neq m; \nonumber \\
\beta_{n m} (-\infty,\infty) \approx &\ \varepsilon \sqrt{2 \pi}\ \mathscr{F} [\Delta \hat{\beta}_{n m}] (\omega^0_n + \omega^0_m). \label{fourier_beta}
\end{align}
That is, the Bogoliubov coefficients between the asymptotic past and future are proportional to the Fourier transforms evaluated at the corresponding substraction (respectively addition) of frequencies. Evidently, resonances occur if the frequency spectrum is peaked around one or more of these values.

Let us remark that, in the presence of resonances, the physically meaningful modes for which the effects take place can be taken as the stationary modes given by~$\{\Psi^0_n \exp (\mp\rmi \omega^0_n t)\}$. This is because the exact modes $\{\prescript{\pm}{}\Phi^{[\tilde{t}]}_n (t)\}$ differ from them just to order~$\varepsilon$ (see~\ref{props_bases}), which is the degree of indefiniteness of the well-defined frequency modes due to the perturbation. Only when the neat effect overcomes this order (and thus the degree of indefiniteness), the effect can be interpreted physically. We shall also mention that the resonance can be consistently described in the regime of duration of the perturbation~$\Delta \tilde{t}$ such that~$1 \ll \omega_{\mathrm{p}} \Delta \tilde{t} \ll 1/\varepsilon$. The reason is that one needs the period of time to be reasonably larger than the inverse of the frequency being described, but on the other hand, one should keep the second order term in~$\varepsilon$ that we dropped in~(\ref{perturbation_Q}) significantly smaller than the first order term that we kept. We refer to Sect.~4 and Appendix~E.2 in Part~I for additional discussion on the interpretation of the modes and on the regime of validity of the perturbative computation.

Finally, we provide the explicit expressions for the entries of~$\Delta K(\tilde{t})$. They are computed in detail in \ref{simplif}. In order to find resonances, the following expressions can be used
\begin{align}
\Delta \hat{\alpha}_{n m} (\tilde{t}) \equiv &\ \rmi \int_{\Sigma^0} \rmd V^0\ [\prescript{-}{m}{\hat{\Delta}}(\tilde{t}) \Psi^0_n] \Psi^0_m \nonumber\\
& + \rmi \int_{\partial \Sigma^0} \rmd S^0\ \Delta x(\tilde{t}) \Big[ (\nabla_{h^0} \Psi^0_n) \cdot (\nabla_{h^0} \Psi^0_m) \nonumber\\
& + (\xi R^{h^0} + m^2 - \omega^0_n \omega^0_m) \Psi^0_n \Psi^0_m \Big], \label{alpha_hat}\\
\Delta \hat{\beta}_{n m} (\tilde{t}) \equiv & - \rmi \int_{\Sigma^0} \rmd V^0\ [\prescript{+}{m}{\hat{\Delta}}(\tilde{t}) \Psi^0_n] \Psi^0_m \nonumber\\
& - \rmi \int_{\partial \Sigma^0} \rmd S^0\ \Delta x(\tilde{t}) \Big[ (\nabla_{h^0} \Psi^0_n) \cdot (\nabla_{h^0} \Psi^0_m) \nonumber\\
& + (\xi R^{h^0} + m^2 + \omega^0_n \omega^0_m) \Psi^0_n \Psi^0_m \Big]; \label{beta_hat}
\end{align}
where~$\rmd S^0$ is the surface element of~$\partial \Sigma^0$, $\nabla_{h^0}$ is the connection associated to the static metric~$h^0_{ij}$,~$R^{h^0}$ its scalar curvature, and~$\prescript{\pm}{m}{\hat{\Delta}}(\tilde{t})$ are linear operators defined by their action on the basis~$\{\Psi^0_n\}$ as
\begin{multline}
\prescript{\pm}{m}{\hat{\Delta}}(\tilde{t}) \Psi^0_n := \\
\big[ \Delta \hat{\mathscr{O}} (\tilde{t}) + \omega^0_n (\omega^0_n \pm \omega^0_m) \Delta r(\tilde{t}) + \xi \Delta \bar{R}(\tilde{t}) \big] \Psi^0_n. \label{superoperator}
\end{multline}

The expressions in (\ref{perturbation_alpha}-\ref{superoperator}) are the second main result of this work. As we will show with concrete examples, they provide a very simple recipe for computing the resonance frequencies and amplitudes of a trapped quantum field in the perturbative regime. We highlight again that they do not even require the computation of the modes to first order in the perturbations, but only the solutions of the static problem in (\ref{op_zeroth}-\ref{psi_zero_norm}).

The symbol~`$\equiv$' in~(\ref{alpha_hat}) and~(\ref{beta_hat}) denotes the equivalence relation ``gives the same resonances as''. This means that the expressions in~(\ref{alpha_hat}) and~(\ref{beta_hat}) have been simplified by dropping terms that are non-zero, but that nonetheless never contribute to the resonances when replaced in~(\ref{perturbation_alpha}) and~(\ref{perturbation_beta}) [or in~(\ref{fourier_alpha}) and~(\ref{fourier_beta})]. Since resonances are the only physically meaningful result to be obtained from this computation, one can always use these expressions to compute the sensibility of the field to each resonance. We refer to \ref{simplif}, and again to Sect.~4 and Appendix~E.2 in Part~I, for more details on the interpretation of the resonances and on the meaning of the ``equivalence for resonances'' relation given by `$\equiv$'.\footnote{This equivalence relation, which concrete expressions are given in \ref{simplif} [equations~(\ref{equiv_res}) and~(\ref{equiv_res_alpha})], can also be used to further simplify concrete expressions for the coefficients found in a specific problem.}

In the expressions~(\ref{alpha_hat}) and~(\ref{beta_hat}) we can see a clear separation between the contributions due to the change of the metric (the volume integrals) and due to the motion of the boundaries (the surface integrals). As it must be the case, the first contributions are equivalent to those found in Part~I [equations~(44) and~(45)].

\subsection{Example: Dynamical Casimir effect}\label{DCE}

In order to provide an illustrative example, let us apply the method to arguably the simplest problem with moving boundary conditions, which is the Dynamical Casimir Effect for a minimally coupled ($\xi = 0$) massive scalar field in $1+1$-dimensional Minkowski spacetime. The spacetime metric is simply
\begin{equation}
\rmd s^2=-\rmd t^2 + \rmd x^2.
\label{1Minkowski}
\end{equation}

The field is trapped inside a cavity of average proper length~$L$, with the boundaries placed at~$x_-$ (left) and~$x_+$ (right). The boundaries oscillate with frequency~$\Omega$ and amplitude $\varepsilon L / 2 \ll L$. We consider three different configurations for such oscillations:
\begin{align}
x_- & = - L/2, \quad x_+ = L[1 + \varepsilon \sin (\Omega t)]/2 \quad & \text{(i)}; \nonumber \\
x_\pm & = \pm L[1 + \varepsilon \sin (\Omega t)]/2 \quad & \text{(ii)}; \label{osc} \\
x_\pm & = \pm L[1 \pm \varepsilon \sin (\Omega t)]/2 \quad & \text{(iii)}. \nonumber
\end{align}
In~(i) only the right boundary oscillates, in~(ii) the boundaries oscillate in opposite directions (the cavity expands and contracts) and in~(iii) the boundaries oscillate in the same direction (the cavity shakes).

For the problem under consideration, it is straightforward to obtain the quantities needed to compute~(\ref{alpha_hat}) and~(\ref{beta_hat}). In particular, we have that
\begin{align}
\Delta \hat{\mathscr{O}} & = \Delta r = \Delta \bar{R} = \prescript{\pm}{m}{\hat{\Delta}} = 0; \nonumber \\
\Delta x (- L/2) & = 0,\ \Delta x (L/2) = L \sin(\Omega t)/2 \ & \text{(i)}; \nonumber \\
\Delta x (\pm L/2) & = L \sin(\Omega t)/2 \ & \text{(ii)}; \label{quantities_dce} \\
\Delta x (\pm L/2) & = \pm L \sin(\Omega t)/2 \ & \text{(iii)}. \nonumber
\end{align}

We solve the problem both for Neumann and Dirichlet boundary conditions (see \ref{dirichlet_app} for the expressions in this latter case). The eigenvalue equation~(\ref{op_zeroth}) and the boundary conditions~(\ref{boundary_zeroth}) [respectively~(\ref{boundary_zeroth_dirichlet})] read
\begin{align}
(-\partial_x^2 + m^2) \Psi^0_n & = (\omega^0_n)^2 \Psi^0_n; \nonumber \\
\pm \partial_x \Psi^0_n|_{x=\pm L/2} & = 0 \quad & \text{(Neumann)}, \label{DCEeigeneq} \\
\Psi^0_n (\pm L/2) & = 0 \quad & \text{(Dirichlet)}; \nonumber
\end{align}
and the solutions to these problems are
\begin{align}
\Psi_n^0 & = \frac{1}{\sqrt{L \omega_n^0}} \cos \left[k_n \left(x + \frac{L}{2}\right)\right] \quad & \text{(Neumann)}, \nonumber \\
\Psi_n^0 & = \frac{1}{\sqrt{L \omega_n^0}} \sin \left[k_n \left(x + \frac{L}{2}\right)\right] \quad & \text{(Dirichlet)}; \label{sols_DCE} \\
\omega_n^0 & = \sqrt{k_n^2 + m^2}; \qquad n \in \mathds{N}^{(*)}; \nonumber
\end{align}
where $k_n := \pi n/L$ and the mode with~$n=0$ is excluded for Dirichlet boundary conditions.

From~(\ref{quantities_dce}) it is immediate that the first integral of the quantities~(\ref{alpha_hat}) and~(\ref{beta_hat}) [respectively of~(\ref{ssd_alpha_hat_moving}) and~(\ref{ssd_beta_hat_moving})] vanishes. Since we are considering one spatial dimension, the ``surface integral'' is simply the evaluation of the integrand at the two boundaries. Plugging the corresponding quantities into~(\ref{alpha_hat}) and~(\ref{beta_hat}), we easily obtain the solutions for Neumann boundary conditions:\footnote{Since we do not need to explicitly consider the evolution in time~$t$ of a basis of modes anymore, we can relax the notation and replace~$\tilde{t} \to t$.}
\begin{equation}
\Delta \hat{\alpha}_{n m} (t) \equiv - \Delta \hat{\beta}_{n m} (t) \equiv \frac{\rmi C_{n m} (\Omega^2 - k_n^2 - k_m^2)}{4 \sqrt{\omega_n^0\omega_m^0}} \sin(\Omega t). \label{hat_DCE}
\end{equation}
The factor~$C_{n m}$ depends on the oscillation configuration, and is given by
\begin{align}
C_{n m} & = (-1)^{n + m} \quad & \text{(i)}, \nonumber \\
C_{n m} & = (-1)^{n + m} + 1 \quad & \text{(ii)}, \label{Cnell} \\
C_{n m} & = (-1)^{n + m} - 1 \quad & \text{(iii)}. \nonumber
\end{align}
In order to obtain the expression in~(\ref{hat_DCE}) we replaced $|\omega_n^0 - \omega_m^0| \to \Omega$ in the computation of~$\Delta \hat{\alpha}_{n m} (t)$ and $\omega_n^0 + \omega_m^0 \to \Omega$ in the computation of~$\Delta \hat{\beta}_{n m} (t)$. This is legitimate within the equivalence relation with respect to resonances, since the only (positive) frequency present in the perturbation is~$\Omega$. Respectively, the solutions for Dirichlet boundary conditions are obtained by plugging the corresponding quantities into~(\ref{ssd_alpha_hat_moving}) and~(\ref{ssd_beta_hat_moving}):
\begin{equation}
\Delta \hat{\alpha}_{n m} (t) \equiv - \Delta \hat{\beta}_{n m} (t) \equiv - \frac{\rmi C_{n m} k_n k_m}{2 \sqrt{\omega_n^0\omega_m^0}} \sin(\Omega t). \label{hat_DCE_dirichlet}
\end{equation}

In general, we find mode mixing and/or particle production due to the moving boundaries, which reproduces the Dynamical Casimir Effect. For example, if the frequency~$\Omega$ coincides with the difference between the frequencies $|\omega^0_n - \omega^0_m|$, plugging~(\ref{hat_DCE}) into~(\ref{perturbation_alpha}) we find that
\begin{equation}
\alpha_{n m} (t_{\mathrm{f}}, t_0) \approx \pm \varepsilon \frac{C_{n m} (\Omega^2 - k_n^2 - k_m^2)}{8 \sqrt{\omega_n^0\omega_m^0}} (t_{\mathrm{f}} - t_0).
\label{alpha_DCE}
\end{equation}
For long enough times, this quantity can overcome the order~$\varepsilon$ and become significant to zeroth order; that is, to the resonant modes of the cavity, and therefore significant mode mixing takes place between the corresponding modes. Analogous arguments apply for the $\beta$-coefficients and the corresponding particle creation, for any boundary conditions and configuration.

We notice that, out of the final results, we can take the limit of a massless field in a straightforward well-defined way.\footnote{The existing ``zero-frequency mode'' in the massless limit (in the case of Neumann boundary conditions) would not be normalisable, but the perturbation does not introduce any effect for it (all the coefficients would vanish for $n=0$ or $m=0$). Thus such mode can simply be ignored.} The results with Dirichlet boundary conditions for configuration~(i) in the massless case exactly reproduce the results obtained in~\cite{PhysRevA.56.4440} and independently in~\cite{Sabin2014}, while for configuration~(iii) they reproduce the results obtained in~\cite{Jorma2013} both for the massive and the massless case.

\subsection{Example: gravitational wave resonance}\label{gw}

Confined quantum fields undergo Bogoliubov transformations when perturbed by gravitational waves. This was shown in~\cite{Sabin2014} considering a scalar field in a one-dimensional rigid trap. The authors proposed to exploit this effect in order to detect gravitational waves using phonons in a Bose-Einstein condensate. In Part~I, we extended this work by computing the field transformations in the three-dimensional case considering free-falling boundary conditions (and thus static in the synchronous gauge). Free-falling boundary conditions were also studied in~\cite{Robbins2018} using a different technique. Considering free-falling boundary conditions is interesting from a mathematical point of view. However, in practice, phononic gravitational wave detectors require inter-atomic interactions and thus, rigid or semi-rigid boundary conditions. The method introduced in this article enables the study of the phonon field transformations in a three-dimensional rigid or semi-rigid cavity. Therefore, the method will be useful in extending~\cite{Sabin2014} to improve the detection of gravitational waves by using three-dimensional trapped Bose-Einstein condensates.

In particular, in this Subsection we explicitly compute the Bogoliubov transformations for the phonon field when trapped in a fully rigid three-dimensional cavity. The phonon field can be described by a real scalar massless quantum field. In the case that the condensate remains stationary, the quantum field obeys a Klein-Gordon equation in an effective metric (with minimal coupling) which corresponds to the gravitational wave metric with the speed of sound in the condensate~$c_\mathrm{s}$ replacing the speed of light in the~$g_{00}$ component~\cite{Visser:2010xv, fagnocchi2010relativistic, bruschi2014testing, hartley2018analogue}. We work in the TT-gauge and normalise the speed of sound $c_\mathrm{s}=1$. We consider a wave with amplitude~$\varepsilon$ and frequency~$\Omega$ propagating in the $z$-direction and with polarisation in the $xy$-directions. Therefore, the metric is given by
\begin{equation}
\rmd s^2 = - \rmd t^2 + [1 + \varepsilon \sin (\Omega t)] \rmd x^2 + [1 - \varepsilon \sin (\Omega t)] \rmd y^2 + \rmd z^2,
\label{metric_gw}
\end{equation}
where we have simplified~$\sin [\Omega (t - z/c)] \to \sin (\Omega t)$, as we have that $c \gg \Omega L_z$ (being~$L_z$ the size of the condensate in the $z$-direction), because of the orders of magnitude between the speed of light and the speed of sound.

For simplicity, we consider that the field is trapped in a rectangular prism of proper lengths~$L_x$, $L_y$ and~$L_z$ aligned with the directions of propagation and polarisation of the wave. Since the cavity is rigid, these proper lengths must stay constant at all times. We consider that the centre of mass of the cavity (which by symmetry coincides with its geometrical centre) is in free-fall, and we fix it at the origin of the coordinate system. Therefore, the boundaries of the prism are placed at (in obvious notation):
\begin{align}
x_{\pm} & = \pm \frac{L_x}{2\sqrt{1+\varepsilon \sin(\Omega t)}}, \nonumber \\
y_{\pm} & = \pm \frac{L_y}{2\sqrt{1-\varepsilon \sin(\Omega t)}}, \label{boundaries_gw} \\
z_{\pm} & = \pm \frac{L_z}{2}. \nonumber
\end{align}

Although we are considering the physical problem of a massless field, we can take advantage of the versatility of our method and address the more general mathematical problem of a massive field with equal ease. The physical problem is then recovered by taking the massless limit. Therefore, from here on we consider~$m \geq 0$. Thus, the eigenvalue equation~(\ref{op_zeroth}) reads
\begin{equation}
(-\partial_x^2 -\partial_y^2 -\partial_z^2 + m^2) \Psi^0_{n m \ell} = (\omega^0_{n m \ell})^2 \Psi^0_{n m \ell};
\label{eigeneq_gw}
\end{equation}
where~$n$, $m$ and~$\ell$ are quantum numbers. We first consider Dirichlet boundary conditions. The boundary conditions imposed to the static modes~(\ref{boundary_zeroth_dirichlet}) are $\Psi^0_{n m \ell} = 0$ at the boundaries given in~(\ref{boundaries_gw}) for $\varepsilon = 0$. The solutions to this problem with the orthonormalisation in~(\ref{psi_zero_norm}) are
\begin{align}
\Psi^0_{n m \ell} = &\ \frac{2}{\sqrt{L_x L_y L_z \omega^0_{n m \ell}}} \sin \left[k^x_n \left(x + \frac{L_x}{2}\right)\right] \nonumber \\
& \times \sin \left[k^y_n \left(y + \frac{L_y}{2}\right)\right]\ \sin \left[k^z_n \left(z + \frac{L_z}{2}\right)\right], \nonumber\\
\omega^0_{n m \ell} = &\ \sqrt{(k^x_n)^2 + (k^y_m)^2 + (k^z_\ell)^2 + m^2}, \quad n, m, \ell \in \mathds{N}^*;
\label{sols_gw}
\end{align}
where~$k^x_n := \pi n / L_x$, and equivalently for the other dimensions. The remaining quantities needed to compute~(\ref{ssd_alpha_hat_moving}) and~(\ref{ssd_beta_hat_moving}) are
\begin{align}
\Delta r & = \Delta \bar{R} = 0, \nonumber \\
\prescript{\pm}{m}{\hat{\Delta}} & = \Delta \hat{\mathscr{O}} = \sin(\Omega t) (\partial_x^2 - \partial_y^2); \nonumber \\
\Delta x (x & = \pm L_x/2) = - L_x \sin (\Omega t) / 4, \label{quantities_gw} \\
\Delta x (y & = \pm L_y/2) = L_y \sin (\Omega t) / 4, \nonumber \\
\Delta x (z & = \pm L_z/2) = 0. \nonumber
\end{align}

Plugging all the quantities into~(\ref{ssd_alpha_hat_moving}) and~(\ref{ssd_beta_hat_moving}), we obtain
\begin{align}
\Delta\hat{\alpha}_{n m \ell}^{n' m' \ell'}(t) \equiv &\ \frac{\rmi\sin(\Omega t)}{4\sqrt{\omega_{n m \ell}^0\omega_{n' m' \ell '}^0}} \delta_\ell^{\ell '} \nonumber\\
& \times \left\{[(-1)^{n+n'}+1]k_n^x k_{n'}^x\delta_m^{m'} \right. \nonumber \\
& \left. - [(-1)^{m+m'}+1]k_m^y k_{m'}^y\delta_n^{n'}\right\},
\label{gw_alpha_hat_moving} \\
\Delta\hat{\beta}_{n m \ell}^{n' m' \ell'}(t) \equiv &\ \frac{\rmi\sin(\Omega t)}{2\omega_{n m \ell}^0} \left[(k_n^x)^2 - (k_m^y)^2\right] \delta_{n m \ell}^{n' m' \ell'} \nonumber\\
& - \Delta\hat{\alpha}_{n m \ell}^{n' m' \ell'}(t).
\label{gw_beta_hat_moving}
\end{align}

An equivalent procedure for Neumann boundary conditions yields the following results:
\begin{align}
\Delta\hat{\alpha}_{n m \ell}^{n' m' \ell'} & (t) \equiv \frac{\rmi\sin(\Omega t)}{8\sqrt{\omega_{n m \ell}^0\omega_{n' m' \ell '}^0}} \delta_\ell^{\ell '} \nonumber\\
& \times \left\{[(-1)^{m+m'}+1][\Omega^2 - (k_m^y)^2 - (k_{m'}^y)^2]\delta_n^{n'} \right. \nonumber \\
& \left. - [(-1)^{n+n'}+1][\Omega^2 - (k_n^x)^2 - (k_{n'}^x)^2]\delta_m^{m'}\right\},
\label{gw_alpha_dirichlet} \\
\Delta\hat{\beta}_{n m \ell}^{n' m' \ell'} & (t) \equiv \frac{\rmi\sin(\Omega t)}{2\omega_{n m \ell}^0} \left[(k_n^x)^2 - (k_m^y)^2\right] \delta_{n m \ell}^{n' m' \ell'} \nonumber\\
& - \Delta\hat{\alpha}_{n m \ell}^{n' m' \ell'}(t);
\label{gw_beta_dirichlet}
\end{align}
where in this case the quantum numbers can take zero values.\footnote{As for the previous example, we have replaced $\tilde{t} \to t$ in the notation, and $|\omega_n^0 \pm \omega_m^0| \to \Omega$ according to the equivalence relation with respect to resonances.} Just as we did in the previous example in Subsect.~\ref{DCE}, we can use~(\ref{perturbation_alpha}) and~(\ref{perturbation_beta}) to compute the linear growing in time of the Bogoliubov coefficients when resonances are present.

Let us give some physical interpretation to the results. The first term in~(\ref{gw_beta_hat_moving}) and~(\ref{gw_beta_dirichlet}) corresponds to the contribution of the perturbation of the metric, and coincides with the result in Subsect.~4.1 of Part~I for free-falling boundaries. The remaining contributions are due to the rigidity of the cavity, and therefore the motion of its boundaries in the TT-gauge. A direct comparison with the results in Subsect.~\ref{DCE} clearly shows that these contributions correspond to a superposition of two ``anti-synchronised'' Dynamical Casimir Effects in the two transversal directions, with configuration~(ii) in the notation of Subsect.~\ref{DCE} (with the cavity expanding and contracting). This is exactly the effect that one would expect from a gravitational wave on a rigid cavity, considering the concrete shape of the cavity and its interaction with the wave.

Thanks to the contributions due to the rigidity of the cavity, both mode-mixing and particle creation between different modes are present. We notice that this could not physically happen in the case of free-falling boundaries. The reason is that any mode-mixing or particle creation between \emph{different} modes (non-diagonal Bogoliubov coefficients) implies local exchange of momentum with the field (in the basis of stationary modes that we are considering). However, a gravitational wave can provide momentum locally to a free field or to a free-falling boundary only in the direction of its propagation, something which in this case is negligible due to the orders of magnitude between the speed of light and the speed of sound. Hence, the direct effect due to the perturbation of the metric is pure cosmological particle creation, which does not exchange momentum and therefore can only affect the diagonal coefficients. On the contrary, the forces keeping the rigidity of the cavity redistribute the momentum so that it is locally non-zero (the boundaries move), and then transmit this momentum locally to the field (although of course the total momentum still vanishes).

Finally, and more interestingly, one can check that for a rigid cavity it is the diagonal quantities~$\Delta\hat{\beta}_{n m \ell}^{n m \ell}(t)$ (the only non-zero quantities for free-falling boundaries) that vanish, for both boundary conditions. This means that the contribution due to the rigidity of the cavity exactly cancels the direct contribution from the change in the metric: Somehow the fact that the cavity keeps its own proper lengths shields the sensibility of the field to any length contractions and expansions from the metric. This is a novel and physically very plausible result. Nonetheless, we think that it is also a non-trivial result, which would be worthy exploring beyond the perturbative regime.

\section{Summary and conclusions}\label{conclu}

In this second article we have extended the method developed in Part~I for computing the evolution of a confined quantum scalar field in a globally hyperbolic spacetime, to the cases in which the timelike boundaries of the spacetime do not remain static in any synchronous gauge. Despite the more sophisticated technical construction required, we have shown that the core ideas of the method can still be used in such situation. Namely, we could construct bases of modes associated to different Cauchy hypersurfaces, a time-dependent linear transformation between them, and a first-order differential equation in time for such transformation. In this case, the coefficients of the transformation depend on the initial conditions of some auxiliary bases, that are solutions to an eigenvalue problem for which the time is just a parameter. If the time-dependent linear transformation connects two regions in which (thanks to a time symmetry) a valid Fock quantisation in terms of the bases of modes associated to each region is possible, then the linear transformation is actually a Bogoliubov transformation, and can be interpreted physically as such in terms of mode-mixing and particle creation between the different modes.

The extension of the method presented here is still of general applicability (as in Part~I), just under some minor assumptions introduced in Subsect.~\ref{statement}. However, we shall stress again that it proves to be especially useful to compute quantitative results on resonances in the perturbative regime (of the metric and the motion of the boundaries). Such usefulness stands out from the simple and practical expressions obtained in Sect.~\ref{res} (and at the end of \ref{dirichlet_app}). We have also illustrated this fact with two examples within the perturbative regime which we could easily solve, namely the Dynamical Casimir Effect (where we reproduced and extended known results) and the perturbation of a field in a rigid cavity by a gravitational wave (which is a completely novel computation). We highlight how the simple expressions obtained [results (\ref{hat_DCE}-\ref{hat_DCE_dirichlet}) and (\ref{gw_alpha_hat_moving}-\ref{gw_beta_dirichlet}), respectively] embrace several physical configurations in an unified way. The perturbative method could also prove its utility in other problems which are now under study, in which quantum systems are perturbed by small gravitational effects \cite{Howl2018, Ratzel2017testing, Ratzel2018dynamical}.

The main aim of this work, both of Part~I and Part~II, is to provide an useful method to compute the evolution of confined quantum fields in concrete physical situations. By applying the method to many different concrete physical problems, mainly (but not only) in the perturbative regime, we have provided plenty of evidence that the method is truly successful in this practical purpose. Specifically, with the examples provided in Part~I and Part~II we reproduce previous results in \cite{Robbins2018,Bernard1977,BD1984,PhysRevA.56.4440,Sabin2014,Jorma2013}. Those results were found using very different approaches and techniques in each work, which implied longer and way more involved calculations. The method presented here manages to reproduce all of the results in an unified way and with a concise calculation for each case. Moreover, the method also extends some of those previous results, easily handling generalisations and variations of them; in particular, some non-trivial variations such as the rigid cavity under a gravitational wave perturbation considered in Subsect.~\ref{gw}. Finally, the results obtained always had consistent physical interpretations. The method will surely prove fruitful in addressing many other relevant problems, and we expect it to become standard in the toolbox of Quantum Field Theory in Curved Spacetime for confined fields, especially in the perturbative regime.

Together with the promising practical applications of the method, there are also future directions of research on the theoretical side. In particular, these include the possible physical interpretations of the time-dependent linear transformations obtained, when they cannot be interpreted directly as Bogoliubov transformations between different Fock quantisations; and their possible connection to field-related (instead of particle-related) quantities. Further extensions of the method for different boundary conditions, quantum fields and/or metric gauges may be also approached.

\begin{acknowledgements}
The authors especially want to thank Stefan Fredenhagen for providing us with the proof given in \ref{proof_geom}, and Jorma Louko for the rich exchange of ideas with us, which greatly helped solving many issues and significantly improved the article. We also want to thank Benito A.\ Ju\'arez-Aubry, Eugenia Colafranceschi, David E.\ Bruschi, Tupac Bravo, Daniel Hartley, Maximilian P.~E.\ Lock, Richard Howl, Joel Lindkvist, Jan Kohlrus, Dennis R\"atzel, Carlos Barcel\'o and Stephan Huimann for their useful comments and discussions during the elaboration of this article. We are also grateful to Gerald Teschl, Miguel S\'anchez Caja, Felix Finster and Simone Murro for clarifying our doubts on the mathematical background. Finally, we would like to thank an anonymous referee for his/her constructive comments, which helped clearing up the expositions of important points of the article. L.~C.~B.\ acknowledges the support from the research platform TURIS, from the \"{O}AW through the project ``Quantum Reference Frames for Quantum Fields'' (ref.~IF\textunderscore 2019\textunderscore 59\textunderscore QRFQF), from the European Commission via Testing the Large-Scale Limit of Quantum Mechanics (TEQ) (No. 766900) project, from the Austrian-Serbian bilateral scientific cooperation no.\ 451-03-02141/2017-09/02, and from the Austrian Science Fund (FWF) through the SFB project BeyondC (sub-project~F7103) and a grant from the Foundational Questions Institute (FQXi) Fund. A.~L.~B.\ recognises support from CONACyT ref:579920/410674. I.~F.\ would like to acknowledge that this project was made possible with the support of the Penrose Institute, the grant ``Quantum Observers in a Relativistic World'' from FQXi's Physics of the Observer program and the grant ``Leaps in cosmology: gravitational wave detection with quantum systems'' (No.~58745) from the John Templeton Foundation. The opinions expressed in this publication are those of the authors and do not necessarily reflect the views of the John Templeton Foundation.
\end{acknowledgements}

\appendix

\section{Why the procedure considered in Part~I does not work here?}\label{why_part_II}

In Part~I, we constructed the modes~$\Phi^{[\tilde{t}]}_n(t)$ by imposing the following initial conditions:
\begin{align}
\hat{\mathscr{O}}(\tilde{t}) \Phi^{[\tilde{t}]}_n (\tilde{t}) & = (\omega^{[\tilde{t}]}_n)^2 \Phi^{[\tilde{t}]}_n (\tilde{t}),
\label{eigenvalues_part_I} \\
\left. \partial_t \Phi^{[\tilde{t}]}_n(t) \right|_{t=\tilde{t}} & = -\rmi \omega^{[\tilde{t}]}_n \Phi^{[\tilde{t}]}_n(\tilde{t}).
\label{derivative_part_I}
\end{align}
In that case, we could solve the eigenvalue problem~(\ref{eigenvalues_part_I}) because the boundary conditions could be written \emph{separately} for the initial condition~$\Phi(\tilde{t})$ and the partial time derivative~$\partial_t \Phi(t) |_{t=\tilde{t}}$. Moreover, these boundary conditions were homogeneous for both quantities, which allowed us to impose equation~(\ref{derivative_part_I}) consistently. On the contrary, in the case where the boundary conditions are in the form of~(\ref{neumann_initial}) (evaluated at $t=\tilde{t}$), they involve in the same equation \emph{both} the gradient of~$\Phi(\tilde{t})$ and the partial time derivative~$\partial_t \Phi(t) |_{t=\tilde{t}}$ (except for the regions of the boundary which remain static). As a consequence, we cannot pose a valid eigenvalue problem just for~$\Phi^{[\tilde{t}]}_n(\tilde{t})$, as we did in Part~I. Therefore, we need to construct the basis of modes associated to each hypersurface~$\Sigma_{\tilde{t}}$ as the solutions to an eigenvalue problem posed directly on the full space of initial conditions $(\Phi(\tilde{t}),\partial_t \Phi(t) |_{t=\tilde{t}})$, so that the boundary conditions for the problem can be properly imposed.

A way to summarise equations~(\ref{eigenvalues_part_I}) and~(\ref{derivative_part_I}) as an eigenvalue problem in the space of initial conditions would be to use the vector eigenvalue equation
\begin{equation}
\left( \begin{array}{cc}
0	& ~\rmi~ \\
	-\rmi \hat{\mathscr{O}}(\tilde{t}) & ~0~
\end{array} \right)
\left( \begin{array}{c}
	\Phi^{[\tilde{t}]}_n (\tilde{t}) \\
	\partial_t \Phi^{[\tilde{t}]}_n (t) |_{t=\tilde{t}}
\end{array} \right)
=
\omega^{[\tilde{t}]}_n
\left( \begin{array}{c}
	\Phi^{[\tilde{t}]}_n (\tilde{t}) \\
	\partial_t \Phi^{[\tilde{t}]}_n (t) |_{t=\tilde{t}}
\end{array} \right).
\label{eigenvalues_bad}
\end{equation}
However, for moving boundary conditions in general the operator in this eigenvalue equation is not self-adjoint. In fact, it is straightforward to find simple problems with moving boundary conditions for which~(\ref{eigenvalues_bad}) has no solutions with real~$\omega^{[\tilde{t}]}_n$. In conclusion: In general, modes with ``locally well-defined frequency'' with respect to time~$t$ [which is how we may call the modes satisfying~(\ref{derivative_part_I})] cannot be solutions to the problem because they do not even fulfil the boundary conditions when the boundaries are moving. This is to be expected, since modes satisfying~(\ref{derivative_part_I}) would be ``locally stationary'' oscillations in phase with respect to~$t$ around~$t=\tilde{t}$, and therefore could not readjust to any motion of the boundary to first order in~$t$.

In order to break this impasse we have considered a Wick rotation in the coordinate time. This rotation has transformed equation~(\ref{eigenvalues_bad}) into the eigenvalue equation~(\ref{eigenvalues_good}), which always provides valid bases of initial conditions, although they do not correspond to modes with ``locally well-defined frequency''.\footnote{The fact that we are immersed in the Wick rotation while manipulating the~$\Psi^{[\tilde{t}]}_n$ modes, is what forces us to use real eigenvectors for their initial conditions.} This Wick rotation is then reversed by the linear transformation~(\ref{psi_to_phi}) from the modes~$\Psi^{[\tilde{t}]}_n$ to the modes~$\prescript{\pm}{}\Phi^{[\tilde{t}]}_n$. Neither these modes can be directly interpreted as modes with ``locally well-defined frequency'' in general. However, this is completely irrelevant, since in the regions~$S$ described in Property~II they do behave as modes with well-defined frequency; and this is all we really need, since those are the only regions where the modes can be used for quantisation. Analogous arguments apply for the case of Dirichlet vanishing boundary conditions as developed in \ref{dirichlet_app}.

Finally, let us notice that, since we are working with different bases of modes as compared to Part~I, the time-dependent linear transformation~(\ref{U_moving_solution}) does not even formally satisfy any Bogoliubov identities. This is because, in general, the bases related are not orthonormal in the Klein-Gordon scalar product. However, the cases in which the transformation can be interpreted as a Bogoliubov transformation are those in which (according to Property~II) the bases related are indeed orthonormal. Therefore, in such cases the coefficients of the transformation do satisfy the Bogoliubov identities.

\section{Proof of the fulfilment of Properties~I and~II}\label{props_bases}

Let us check first Property~I, namely that the set $\{\prescript{\pm}{}\Phi^{[\tilde{t}]}_n\}$ is a basis of the space of solutions to the Klein-Gordon equation. We first realise that it is equivalent to check that the set $\{\Psi^{[\tilde{t}]}_n\} = \{\prescript{\pm}{}\Psi^{[\tilde{t}]}_n\}$ is also a basis of the space of solutions, since the set~$\{\prescript{\pm}{}\Phi^{[\tilde{t}]}_n\}$ is obtained by the linear invertible transformation~(\ref{psi_to_phi}) of the elements in~$\{\prescript{\pm}{}\Psi^{[\tilde{t}]}_n\}$. Because of Condition~B and the linearity of the Klein-Gordon equation, this is the case if and only if the set of initial conditions $\{(\Psi^{[\tilde{t}]}_n (\tilde{t}), \partial_t \Psi^{[\tilde{t}]}_n (t)|_{t=\tilde{t}})\}$ is a basis of the space of initial conditions at~$\Sigma_{\tilde{t}}$, which is clearly the space~$L^2(\Sigma_{\tilde{t}}) \oplus L^2(\Sigma_{\tilde{t}})$. This is definitely fulfilled, since this set of initial conditions is constructed with the eigenvectors of the self-adjoint operator~$\hat{\mathscr{M}} (\tilde{t})$, that are solutions to the eigenvalue problem~(\ref{eigenvalues_good}).

In order to check Property~II, let us consider a spacetime region~$S$ in which the conditions listed in Property~II hold. Then, it is easy to check that, for each mode~$\prescript{+}{}\Psi^{[\tilde{t}]}_n (t)$ with positive eigenvalue $\prescript{+}{}\omega^{[\tilde{t}]}_n > 0$ there is a corresponding mode~$\prescript{-}{}\Psi^{[\tilde{t}]}_n (t)$ with negative eigenvalue $\prescript{-}{}\omega^{[\tilde{t}]}_n = - (\prescript{+}{}\omega^{[\tilde{t}]}_n)$, and with the same initial condition~$\prescript{-}{}\Psi^{[\tilde{t}]}_n (\tilde{t}) = \prescript{+}{}\Psi^{[\tilde{t}]}_n (\tilde{t})$. If we put each mode of the pair in one of the subsets~$\{\prescript{\pm}{}\Psi^{[\tilde{t}]}_n\}$ and we use the same index~$n$ for them, then the modes~$\prescript{\pm}{}\Phi^{[\tilde{t}]}_n$ satisfy the following initial conditions in~$\Sigma_{\tilde{t}}$:
\begin{align}
\prescript{\pm}{}\Phi^{[\tilde{t}]}_n (\tilde{t}) & = \prescript{+}{}\Psi^{[\tilde{t}]}_n (\tilde{t}), \nonumber\\
\partial_t \prescript{\pm}{}\Phi^{[\tilde{t}]}_n(t) |_{t=\tilde{t}} & = \mp \rmi (\prescript{+}{}\omega^{[\tilde{t}]}_n) \prescript{+}{}\Psi^{[\tilde{t}]}_n(\tilde{t}). \label{recovering_zero}
\end{align}
Moreover, the eigenvalue problem~(\ref{eigenvalues_good}) has the same solutions for all the hypersurfaces~$\Sigma_t$ within~$S$. Taking into account these facts, and also that we are considering only real spatial functions~$\prescript{\pm}{}\Psi^{[\tilde{t}]}_n(\tilde{t})$, the modes of the form~(\ref{modes_with_freq}) satisfy both the initial conditions~(\ref{recovering_zero}) in~$\Sigma_{\tilde{t}}$ and the Klein-Gordon equation~(\ref{klein-gordon_2}) in the whole region~$S$. Therefore, these modes correspond to the modes that we are actually assigning to the hypersurface~$\Sigma_{\tilde{t}}$. Now, if the interval of time that~$S$ embraces is large enough so as to explore the minimum frequency in the spectrum, then we can talk about the modes~(\ref{modes_with_freq}) as modes with well-defined positive and negative frequency with respect to~$t$.

In Subsect.~\ref{bases} we claimed that, for the problems in which we can give a physical interpretation to the results (that is, the problems where Fock quantisation is possible at least in some regions), there is an infinite number of both positive and negative eigenvalues~$\omega^{[\tilde{t}]}_n$ for every~$\tilde{t}$. Now we can justify this claim. The reason is that, for regular metrics, the eigenvalues obtained at the different hypersurfaces~$\Sigma_{\tilde{t}}$ should be a continuous function of~$\tilde{t}$. Since, at least in the regions where Fock quantisation is possible, these eigenvalues are indeed divided into infinitely many positive and negative, and since in no case $\omega^{[\tilde{t}]}_n = 0$ (see \ref{no_zeros}), then they must stay divided in such way even in the regions where Fock quantisation is not possible.

It remains to be proven that, under the conditions given in Property~II for the region~$S$, the basis~$\{\prescript{\pm}{}\Phi^{[\tilde{t}]}_n\}$ is orthonormal. First, let us write down the Klein-Gordon inner product~(\ref{scalar_product}) between the modes~$\prescript{\pm}{}\Psi^{[\tilde{t}]}_n$. Using~(\ref{first_derivative}) we get
\begin{multline}
\langle \prescript{\pm}{}\Psi^{[\tilde{t}]}_n, \prescript{\hat{\pm}}{}\Psi^{[\tilde{t}]}_m \rangle = \\
\rmi \left[ (\prescript{\pm}{}\omega^{[\tilde{t}]}_n) - (\prescript{\hat{\pm}}{}\omega^{[\tilde{t}]}_m) \right] \int_{\Sigma_{\tilde{t}}} \rmd V_{\tilde{t}}\ \prescript{\pm}{}\Psi^{[\tilde{t}]}_n (\tilde{t}) \prescript{\hat{\pm}}{}\Psi^{[\tilde{t}]}_m (\tilde{t}).
\label{kg_psi}
\end{multline}

We need to evaluate the integral in~(\ref{kg_psi}). In order to do so, we evaluate the inner product~(\ref{scalar_product_ini}) between the initial conditions of these two modes and compare this evaluation with the orthonormalisation condition that we imposed in~(\ref{norma_casera_psi}). Using~(\ref{eigenvalues_good}) and Green's first identity, we obtain
\begin{multline}
\prescript{\pm}{}\omega^{[\tilde{t}]}_n \left[ (\prescript{\pm}{}\omega^{[\tilde{t}]}_n) + (\prescript{\hat{\pm}}{}\omega^{[\tilde{t}]}_m) \right] \int_{\Sigma_{\tilde{t}}} \rmd V_{\tilde{t}}\ \prescript{\pm}{}\Psi^{[\tilde{t}]}_n (\tilde{t}) \prescript{\hat{\pm}}{}\Psi^{[\tilde{t}]}_m (\tilde{t}) \\
+ \int_{\partial \Sigma_{\tilde{t}}}\rmd S_{\tilde{t}} \prescript{\hat{\pm}}{}\Psi^{[\tilde{t}]}_m (\tilde{t})\ \vec{n} \cdot \nabla_{h(\tilde{t})} \prescript{\pm}{}\Psi^{[\tilde{t}]}_n (\tilde{t}) = |\prescript{\pm}{}\omega^{[\tilde{t}]}_n| \delta_{nm} \delta_{\pm \hat{\pm}}.
\label{compare_sp}
\end{multline}
Equation~(\ref{compare_sp}) is general, that is, we have not used yet the specific conditions holding in region~$S$. Let us now impose those conditions. In particular, imposing~$v_{\mathrm{B}} (t) = 0$ and using~(\ref{neumann_moving_modes_boundary}) we have that the surface integral vanishes. If we pick now the~`$++$' sign prescripts, noticing that $\prescript{+}{}\omega^{[\tilde{t}]}_n > 0$ we obtain
\begin{equation}
\int_{\Sigma_{\tilde{t}}} \rmd V_{\tilde{t}}\ \prescript{+}{}\Psi^{[\tilde{t}]}_n (\tilde{t}) \prescript{+}{}\Psi^{[\tilde{t}]}_m (\tilde{t}) = \frac{\delta_{n m}}{2 (\prescript{+}{}\omega^{[\tilde{t}]}_n)}.
\label{psi_norm}
\end{equation}
Moreover, we also have that~$\prescript{+}{}\Psi^{[\tilde{t}]}_n (\tilde{t}) = \prescript{-}{}\Psi^{[\tilde{t}]}_m (\tilde{t})$, so the previous equation holds for any sign prescripts of the spatial functions. Using this fact and $\prescript{-}{}\omega^{[\tilde{t}]}_n = - (\prescript{+}{}\omega^{[\tilde{t}]}_n)$, we can finally compute the Klein-Gordon inner product in~(\ref{kg_psi}), obtaining
\begin{equation}
\langle \prescript{\pm}{}\Psi^{[\tilde{t}]}_n, \prescript{\hat{\pm}}{}\Psi^{[\tilde{t}]}_m \rangle = \pm \rmi \delta_{nm} \delta_{\mp \hat{\pm}}.
\label{kg_psi_solved}
\end{equation}
Finally, using the linear transformation~(\ref{psi_to_phi}), we get
\begin{equation}
\langle \prescript{\pm}{}\Phi^{[\tilde{t}]}_n, \prescript{\hat{\pm}}{}\Phi^{[\tilde{t}]}_m \rangle = \pm \delta_{n m} \delta_{\pm \hat{\pm}},
\label{kg_phi}
\end{equation}
which proves that the basis is orthonormal.

\section{Proof of the differential equation for the transformation}\label{great_proof}

Let us call~$V(\tilde{t}, \tilde{t}_0)$ the linear transformation between the basis~$\{\prescript{\pm}{}\Psi^{[\tilde{t}_0]}_n\}$ and the basis~$\{\prescript{\pm}{}\Psi^{[\tilde{t}]}_n\}$. By composition of linear transformations, it is clear that
\begin{equation}
V(\tilde{t} + \Delta \tilde{t}, \tilde{t}_0) = V(\tilde{t} + \Delta \tilde{t}, \tilde{t}) V(\tilde{t}, \tilde{t}_0).
\label{V_composition}
\end{equation}
We can use this relation to obtain the following differential equation for~$V(\tilde{t}, \tilde{t}_0)$:
\begin{align}
\frac{\rmd}{\rmd \tilde{t}} V(\tilde{t}, \tilde{t}_0) &= \left. \frac{\rmd}{\rmd (\Delta \tilde{t})} V(\tilde{t} + \Delta \tilde{t}, \tilde{t}_0) \right|_{\Delta \tilde{t} = 0} \notag \\
&= \left. \frac{\rmd}{\rmd (\Delta \tilde{t})} V(\tilde{t} + \Delta \tilde{t}, \tilde{t}) \right|_{\Delta \tilde{t} = 0} V(\tilde{t}, \tilde{t}_0).
\label{transformation_differentiation}
\end{align}
Notice now that, because of the relation~(\ref{psi_to_phi}) between bases, we have that the linear transformation~$U(\tilde{t}, \tilde{t}_0)$ is related to the linear transformation~$V(\tilde{t}, \tilde{t}_0)$ by
\begin{equation}
U(\tilde{t}, \tilde{t}_0) = M V(\tilde{t}, \tilde{t}_0) M^*,
\label{V_to_U}
\end{equation}
since $M^{-1} = M^*$. Then, if we call~$\hat{V}(\tilde{t})$ the first factor on the r.h.s. of~(\ref{transformation_differentiation}), replacing~(\ref{V_to_U}) we get the differential equation~(\ref{differential_equation_moving_U}). Therefore, what remains to be obtained are the expressions for~$\hat{V}(\tilde{t})$ given by~(\ref{def_Vhat}) and~(\ref{V_hat}) out of the definition
\begin{equation}
\hat{V}(\tilde{t}) := \left. \frac{\rmd}{\rmd (\Delta \tilde{t})} V(\tilde{t} + \Delta \tilde{t}, \tilde{t}_0) \right|_{\Delta \tilde{t} = 0},
\label{V_hat_real_def}
\end{equation}
where~$V(\tilde{t} + \Delta \tilde{t}, \tilde{t})$ is so far only implicitly defined by being the linear transformation between the~$\{\prescript{\pm}{}\Psi^{[\tilde{t}]}_n\}$ bases at different times.

Since the basis~$\{\prescript{\pm}{}\Psi^{[\tilde{t}]}_n\}$ has the corresponding orthonormalisation given by~(\ref{norma_casera_psi}), it is clear that we can compute the elements of~$V(\tilde{t} + \Delta \tilde{t}, \tilde{t})$ as
\begin{equation}
V^{\pm \hat{\pm}}_{n m}(\tilde{t} + \Delta \tilde{t}, \tilde{t}) = \frac{1}{|\prescript{\hat{\pm}}{}\omega_m^{[\tilde{t}]}|}\langle\prescript{\pm}{}\Psi_n^{[\tilde{t}+\Delta\tilde{t}]},\prescript{\hat{\pm}}{}\Psi_m^{[\tilde{t}]}\rangle_{\Sigma_{\tilde{t}}}.
\label{comp_V}
\end{equation}
According to the definition of the inner product between modes associated to the hypersurface~$\Sigma_{\tilde{t}}$, which is given by the inner product in~(\ref{scalar_product_ini}) with the values of the corresponding modes and their first time derivatives at~$\Sigma_{\tilde{t}}$ as entries, we have that
\begin{multline}
\hat{V}^{\pm \hat{\pm}}_{nm}(\tilde{t}) = \frac{1}{|\prescript{\hat{\pm}}{}\omega_m^{[\tilde{t}]}|}\frac{\rmd}{\rmd\Delta\tilde{t}} V^{\pm \hat{\pm}}_{n m}(\tilde{t} + \Delta \tilde{t}, \tilde{t}) = \\
\frac{1}{|\prescript{\hat{\pm}}{}\omega_m^{[\tilde{t}]}|}\frac{\rmd}{\rmd\Delta\tilde{t}}\left\{ \int_{\Sigma_{\tilde{t}}}\rmd V_{\tilde{t}}\left.\partial_{t}\prescript{\pm}{}\Psi_n^{[\tilde{t}+\Delta\tilde{t}]}(t)\right|_{t=\tilde{t}}\left.\partial_{t}\prescript{\hat{\pm}}{}\Psi_m^{[\tilde{t}]}(t)\right|_{t=\tilde{t}} \right. \\
\left.+\int_{\Sigma_{\tilde{t}}}\rmd V_{\tilde{t}}\left[\xi R^{h}(\tilde{t})+m^2 + F(\tilde{t})\right]\prescript{\pm}{}\Psi_n^{[\tilde{t}+\Delta\tilde{t}]}(\tilde{t})\prescript{\hat{\pm}}{}\Psi_m^{[\tilde{t}]}(\tilde{t})\right.\\
\left.+\int_{\Sigma_{\tilde{t}}}\rmd V_{\tilde{t}}\left[\nabla_{h(\tilde{t})}\prescript{\pm}{}\Psi_n^{[\tilde{t}+\Delta\tilde{t}]}(\tilde{t})\right]\cdot\left[\nabla_{h(\tilde{t})}\prescript{\hat{\pm}}{}\Psi_m^{[\tilde{t}]}(\tilde{t})\right]\right\}.
\label{V_hat_explicit}
\end{multline}

We need to compute the derivative with respect to~$\Delta \tilde{t}$ of the quantities inside the integrals. For that, we use the \emph{local} evolution in time around~$\tilde{t} + \Delta \tilde{t}$ (and to first order in~$\Delta \tilde{t}$) of~$\prescript{\pm}{}\Psi^{[\tilde{t} + \Delta \tilde{t}]}_n(t)$, which is given by~(\ref{first_derivative}); and of~$\partial_t \prescript{\pm}{}\Psi^{[\tilde{t} + \Delta \tilde{t}]}_n(t)$, which we obtain through the Klein-Gordon equation. Notice that conditions~(\ref{neumann_moving_modes}) and~(\ref{first_derivative}), when replaced in the Klein-Gordon equation as written in~(\ref{klein-gordon_2}) and evaluated at~$t=\tilde{t}$, imply that
\begin{align}
\left. \partial^2_t \prescript{\pm}{}\Psi^{[\tilde{t}]}_n(t) \right|_{t=\tilde{t}} =\ & - (\prescript{\pm}{}\omega^{[\tilde{t}]}_n) \left[ \prescript{\pm}{}\omega^{[\tilde{t}]}_n + q(\tilde{t}) \right] \prescript{\pm}{}\Psi^{[\tilde{t}]}_n(\tilde{t}) \nonumber \\
& - \left[\xi\bar{R}(\tilde{t})-F(\tilde{t})\right] \prescript{\pm}{}\Psi^{[\tilde{t}]}_n(\tilde{t}).
\label{second_derivative_moving}
\end{align}
That is, out of the initial conditions, the Klein-Gordon equation provides the value of the second time derivative of~$\prescript{\pm}{}\Psi^{[\tilde{t}]}_n(t)$ at~$t=\tilde{t}$ [and \emph{only} at~$t=\tilde{t}$, equation~(\ref{second_derivative_moving}) is evidently \emph{not} a differential equation in time].

Because of relations~(\ref{first_derivative}) and~(\ref{second_derivative_moving}) (replacing $\tilde{t} \to \tilde{t} + \Delta \tilde{t}$), we have that
\begin{multline}
\prescript{\pm}{}\Psi_n^{[\tilde{t}+\Delta\tilde{t}]}(\tilde{t})= \\
\prescript{\pm}{}\Psi_n^{[\tilde{t}+\Delta\tilde{t}]}(\tilde{t}+\Delta\tilde{t})\left[1-(\prescript{\pm}{}\omega_n^{[\tilde{t}]})\Delta\tilde{t}\right] +\mathcal{O}\left(\Delta\tilde{t}\right)^2, \label{expansion_psi}
\end{multline}
\begin{multline}
\left.\partial_{t}\prescript{\pm}{}\Psi_n^{[\tilde{t}+\Delta\tilde{t}]}(t)\right|_{t=\tilde{t}}= \\
\prescript{\pm}{}\omega_n^{[\tilde{t}+\Delta\tilde{t}]}\prescript{\pm}{}\Psi_n^{[\tilde{t}+\Delta\tilde{t}]}(\tilde{t}+\Delta\tilde{t}) \left\{ 1 + \left[\prescript{\pm}{}\omega_n^{[\tilde{t}]}+q(\tilde{t})\right] \Delta\tilde{t} \right\} \\
+\left[\xi\bar{R}(\tilde{t})-F(\tilde{t})\right]\prescript{\pm}{}\Psi_n^{[\tilde{t}]}(\tilde{t})\Delta\tilde{t} +\mathcal{O}\left(\Delta\tilde{t}\right)^2. \label{expansion_dpsi}
\end{multline}

This is the local evolution of the needed quantities to first order in~$\Delta \tilde{t}$, and therefore we are ready to compute the derivative in~(\ref{V_hat_explicit}). Plugging~(\ref{expansion_psi}) and~(\ref{expansion_dpsi}) into~(\ref{V_hat_explicit}), and using Green's first identity, the orthonormalisation condition in~(\ref{norma_casera_psi}) and the boundary condition~(\ref{neumann_moving_modes_boundary}), after a tedious but straightforward calculation we obtain the expression for the elements in~(\ref{V_hat}). This completes the proof of equation~(\ref{differential_equation_moving_U}).

\section{Derivation of the expressions in the perturbative regime}\label{simplif}

The entries~$\Delta \hat{\alpha}_{nm} (\tilde{t})$ and~$\Delta \hat{\beta}_{nm} (\tilde{t})$ in~$\Delta K (\tilde{t})$ (and their respective complex conjugates) correspond to the first order contributions in~$\varepsilon$ to the factor~$M \hat{V} (\tilde{t}) M^*$ in~(\ref{differential_equation_moving_U}). In order to compute them, we temporarily introduce the expressions for the solutions to first order in~$\varepsilon$:\footnote{From here on, we omit the explicit dependence on~$\tilde{t}$ for most of the quantities.}
\begin{align}
\prescript{\pm}{}\Psi^{[\tilde{t}]}_n(\tilde{t}) & \approx \Psi^0_n + \varepsilon \prescript{\pm}{}\Delta \Psi_n, \nonumber \\
\prescript{\pm}{}\omega^{[\tilde{t}]}_n & \approx \pm (\omega^0_n + \varepsilon \prescript{\pm}{}\Delta \omega_n). \label{perturbation_sols}
\end{align}
Just as the solutions to the static problem satisfied the zeroth order in~$\varepsilon$ of~(\ref{neumann_moving_modes}) and~(\ref{neumann_moving_modes_boundary}) [given by~(\ref{op_zeroth}) and~(\ref{boundary_zeroth})], the perturbations appearing in~(\ref{perturbation_sols}) must satisfy the first order in~$\varepsilon$ of those equations, which reads
\begin{align}
& \hat{\mathscr{O}}^0 (\prescript{\pm}{}\Delta \Psi_n) + \Delta\hat{\mathscr{O}} \Psi^0_n = (\omega^0_n)^2 (\prescript{\pm}{}\Delta \Psi_n) + 2 \omega^0_n (\prescript{\pm}{}\Delta \omega_n) \Psi^0_n,
\label{op_first_pm} \\
& \vec{n} \cdot \nabla_{h^0} \prescript{\pm}{}\Delta \Psi_n + \Delta x \vec{n} \cdot \nabla_{h^0} \left(\vec{n} \cdot \nabla_{h^0} \Psi^0_n \right) = \mp \omega^0_n \Psi^0_n \frac{\rmd}{\rmd \tilde{t}} \Delta x;
\label{boundary_first}
\end{align}
where all the quantities in the second equation are evaluated at the boundary~$\partial \Sigma^0$. The second term of the second equation takes into account the displacement of the point where the boundary condition is imposed. We also temporarily introduce the perturbation of $F(\tilde{t})$ to first order in $\varepsilon$ (since we need to prove here that indeed it does not contribute to the resonances):
\begin{equation}
F \approx \varepsilon \Delta F.
\label{Delta_F}
\end{equation}

Let us calculate the first order in~$\varepsilon$ of the quantity~$\hat{V}^{\pm \hat{\pm}}_{nm}$ in~(\ref{V_hat}), which in terms of the perturbed quantities in~(\ref{perturbation_quantities}), (\ref{perturbation_sols}) and~(\ref{Delta_F}) reads
\begin{multline}
\left. \frac{\partial}{\partial \varepsilon} \hat{V}^{\pm \hat{\pm}}_{nm} \right|_{\varepsilon = 0} = \hat{\pm} \Bigg\{ - (\prescript{\pm}{}\Delta \omega_n) \delta_{nm} \delta_{\pm \hat{\pm}} \\
+ (\pm \omega^0_n \hat{\pm} \omega^0_m) \int_{\Sigma^0} \rmd V^0 \left(\frac{\rmd}{\rmd \tilde{t}} \prescript{\pm}{}\Delta \Psi_n \right) \Psi^0_m \\
+ 2 (\prescript{\pm}{}\Delta \omega_n) \delta_{nm} + \left(\pm \frac{\rmd \prescript{\pm}{}\Delta \omega_n}{\rmd \tilde{t}} - \Delta F \right) \frac{\delta_{nm}}{2 \omega^0_n} \\
+ 2 (\omega^0_n)^2 \left[ \int_{\Sigma^0} \rmd V^0 (\prescript{\pm}{}\Delta \Psi_n) \Psi^0_m + \Delta J^{\hat{\pm}}_{nm} \right] \\
+ \int_{\Sigma^0} \rmd V^0\ \Psi^0_n \left(\pm \omega^0_n \Delta q + \xi \Delta \bar{R} \right) \Psi^0_m \Bigg\},
\label{V_hat_perturb}
\end{multline}
where $\Delta q := \partial_\varepsilon q |_{\varepsilon=0}$ and~$\Delta J^\pm_{nm}$ is defined as
\begin{equation}
\Delta J^\pm_{nm} := \left. \frac{\partial}{\partial \varepsilon} \int_{\Sigma_{\tilde{t}}} \rmd V_{\tilde{t}}\ \Psi^0_n \prescript{\pm}{}\Psi^{[\tilde{t}]}_m (\tilde{t}) \right|_{\varepsilon = 0}.
\label{def_J}
\end{equation}
This quantity contains contributions from~$\prescript{\pm}{}\Psi^{[\tilde{t}]}_m (\tilde{t})$, but also from~$\rmd V_{\tilde{t}}$ and from the change in the domain~$\Sigma_{\tilde{t}}$, since these two objects depend on~$\varepsilon$. In \ref{proof_J} we prove that this quantity takes the value
\begin{align}
\Delta J^\pm_{nm} = &\ -\frac{1}{2 (\omega^0_n)^2} \Bigg\{ \frac{1}{2} \left[ (\prescript{+}{}\Delta \omega_n) + (\prescript{-}{}\Delta \omega_n) \right] \delta_{nm} \nonumber\\
&\ + \omega^0_n (\omega^0_n \pm \omega^0_m) \int_{\Sigma^0} \rmd V^0\ \prescript{+}{}\Delta \Psi_n \Psi^0_m \nonumber\\
&\ + \omega^0_n (\omega^0_n \mp \omega^0_m) \int_{\Sigma^0} \rmd V^0\ \prescript{-}{}\Delta \Psi_n \Psi^0_m \Bigg\}.
\label{value_J}
\end{align}

We provide the explicit computation of $\Delta \hat{\beta}_{nm} (\tilde{t})$. The computations of the other entries in~$\Delta K (\tilde{t})$ follow an analogous procedure. Using~(\ref{matrix_M}) and~(\ref{def_Vhat}) we have that
\begin{equation}
\Delta \hat{\beta}_{nm} (\tilde{t}) = \frac{1}{2} \frac{\partial}{\partial \varepsilon} \left[ \hat{V}^{+-}_{nm} + \hat{V}^{-+}_{nm} + \rmi (\hat{V}^{--}_{nm} - \hat{V}^{++}_{nm}) \right]_{\varepsilon = 0}.
\label{beta_hat_perturb_def}
\end{equation}

The first part of the computation consists of plugging the expressions~(\ref{V_hat_perturb}) and~(\ref{value_J}) into~(\ref{beta_hat_perturb_def}). We also use at this point the equivalence relation with respect to resonances for the terms appearing in~$\Delta \hat{\beta}_{nm} (\tilde{t})$, which reads
\begin{equation}
\frac{\rmd X (\tilde{t})}{\rmd \tilde{t}} \equiv \rmi (\omega^0_n + \omega^0_m) X(\tilde{t}),
\label{equiv_res}
\end{equation}
where~$X(\tilde{t})$ can be any function of~$\tilde{t}$. As we advanced, this equivalence means that, although the two quantities are not the same, when~$X(\tilde{t})$ contains a term with the correct resonant frequency $\omega^0_n + \omega^0_m$, the contribution of this term to the Bogoliubov coefficient~(\ref{perturbation_beta}) or~(\ref{fourier_beta}) is the same for the two quantities in~(\ref{equiv_res}). We refer to Appendix~E.2 in Part~I for a careful mathematical derivation of this equivalence. We use the equivalence in order to simplify the derivatives with respect to~$\tilde{t}$ appearing in~(\ref{V_hat_perturb}). After a tedious but mechanical calculation, we obtain
\begin{align}
\Delta \hat{\beta}_{nm} (\tilde{t}) & \equiv - \rmi \left[(\omega^0_n)^2 - (\omega^0_m)^2 \right] \int_{\Sigma^0} \rmd V^0\ \Delta \Psi_n \Psi^0_m \nonumber\\
& - \rmi \int_{\Sigma^0} \rmd V^0\ \Psi^0_n \left[\omega^0_n (\omega^0_n + \omega^0_m) \Delta r + \xi \Delta \bar{R} \right] \Psi^0_m \nonumber\\
& - \rmi \left( \Delta \omega_n - \frac{\Delta F}{2 \omega^0_n} \right) \delta_{nm};
\label{beta_hat_perturb_int}
\end{align}
where $\Delta r$ is given by~(\ref{def_dr}), and we have defined the quantities
\begin{align}
\Delta \Psi_n & := \frac{1}{2} \left[ (1-\rmi) \prescript{+}{}\Delta \Psi_n + (1+\rmi) \prescript{-}{}\Delta \Psi_n \right],
\label{def_delta_psi} \\
\Delta \omega_n & := \frac{1}{2} \left[ (1-\rmi) \prescript{+}{}\Delta \omega_n + (1+\rmi) \prescript{-}{}\Delta \omega_n \right].
\label{def_delta_omega}
\end{align}

The second part of the computation consists of a first attempt for getting rid of the perturbations of the solutions~$\Delta \Psi_n$ and~$\Delta \omega_n$ (usually harder to compute for a given problem) by introducing the perturbation of the operator~$\Delta \hat{\mathscr{O}}$ in~(\ref{perturbation_quantities}) (which is in principle trivial to compute). By considering a linear combination of~(\ref{op_first_pm}) for the positive and negative prescripts, we can write that equation for the quantities defined in~(\ref{def_delta_psi}) and~(\ref{def_delta_omega}):
\begin{equation}
\hat{\mathscr{O}}^0 \Delta \Psi_n + \Delta\hat{\mathscr{O}} \Psi^0_n = (\omega^0_n)^2 \Delta \Psi_n + 2 \omega^0_n \Delta \omega_n \Psi^0_n.
\label{op_first}
\end{equation}
Using (\ref{op_zeroth}) and~(\ref{op_first}), Green's second identity and the known properties of the solutions to zeroth order, we can do the following calculation:
\begin{align}
(\omega^0_m)^2 & \int_{\Sigma^0} \rmd V^0\ \Delta \Psi_n \Psi^0_m = \int_{\Sigma^0} \rmd V^0\ \Delta \Psi_n (\hat{\mathscr{O}}^0 \Psi^0_m) \nonumber\\
= & \int_{\Sigma^0} \rmd V^0 (\hat{\mathscr{O}}^0 \Delta \Psi_n) \Psi^0_m + \int_{\partial \Sigma^0} \rmd S^0\ \Psi^0_m \vec{n} \cdot \nabla_{h^0} \Delta \Psi_n \nonumber\\
= &\ - \int_{\Sigma^0} \rmd V^0 (\Delta \hat{\mathscr{O}} \Psi_n^0) \Psi_m^0 + (\omega^0_n)^2 \int_{\Sigma^0} \rmd V^0\ \Delta \Psi_n \Psi^0_m \nonumber\\
&\ + \Delta \omega_n \delta_{nm} + \int_{\partial \Sigma^0} \rmd S^0\ \Psi^0_m \vec{n} \cdot \nabla_{h^0} \Delta \Psi_n.
\label{supersimpl}
\end{align}
Rearranging terms between the first and the last line we can use this computation to simplify~(\ref{beta_hat_perturb_int}). In particular, the direct contribution of~$\Delta F$ in~(\ref{beta_hat_perturb_int}) cancels out with its contribution through~$\Delta \hat{\mathscr{O}}$. Also, after the simplification the only remaining quantity that could depend on~$\Delta F$ is $\vec{n} \cdot \nabla_{h^0} \Delta \Psi_n$ in the surface integral in~(\ref{supersimpl}). But from~(\ref{boundary_first}) and~(\ref{def_delta_psi}) one can see that such quantity evaluated at the surface is fully determined by other quantities which do not depend on~$\Delta F$. Therefore, we have already proven that~$\Delta F$ does not contribute to the resonances, and from here on we can consider again $F=0$. Consequently, (\ref{beta_hat_perturb_int}) becomes
\begin{align}
\Delta \hat{\beta}_{nm} (\tilde{t}) \equiv &\ - \rmi \int_{\Sigma^0} \rmd V^0\ (\prescript{+}{m}{\hat{\Delta}} \Psi^0_n) \Psi^0_m \nonumber\\
&\ + \rmi \int_{\partial \Sigma^0} \rmd S^0\ \Psi^0_m \vec{n} \cdot \nabla_{h^0} \Delta \Psi_n,
\label{beta_hat_perturb_int_2}
\end{align}
where~$\prescript{+}{m}{\hat{\Delta}} (\tilde{t})$ has been defined in~(\ref{superoperator}).

We can see that the perturbation of the modes still appears in the remaining surface integral. The third and last part of the computation consists of further manipulating this surface integral in order to fully get rid of the perturbation of the modes. In this last part of the computation for convenience we consider the notation simplifications $h^0 \to h$ and $\nabla_{h^0} \to \nabla$. We notice that the boundary condition~(\ref{boundary_first}) for the quantities in~(\ref{def_delta_psi}) and~(\ref{def_delta_omega}) reads
\begin{equation}
\vec{n} \cdot \nabla \Delta \Psi_n + \Delta x \vec{n} \cdot \nabla \left(\vec{n} \cdot \nabla \Psi^0_n \right) = \rmi \omega^0_n \Psi^0_n \frac{\rmd}{\rmd \tilde{t}} \Delta x.
\label{boundary_first_sum}
\end{equation}
We can use this boundary condition and again the equivalence~(\ref{equiv_res}) to write
\begin{equation}
\vec{n} \cdot \nabla \Delta \Psi_n \equiv - \Delta x\ \vec{n} \cdot \nabla \left(\vec{n} \cdot \nabla \Psi^0_n \right) - \omega^0_n (\omega^0_n + \omega^0_m) \Psi^0_n \Delta x.
\label{equiv_bound}
\end{equation}
We split the first term on the r.h.s.\ into two terms using index notation:
\begin{equation}
\vec{n} \cdot \nabla \left(\vec{n} \cdot \nabla \Psi^0_n \right) = (n^i \nabla_i n^j) \nabla_j \Psi^0_n + n^i n^j \nabla_i \nabla_j \Psi^0_n.
\label{split_surf}
\end{equation}
In \ref{proof_geom} we prove the following relation~\cite{stefan}:
\begin{equation}
n^i \nabla_i n^j = \frac{1}{\Delta x} (n^i n^j - h^{i j}) \nabla_i \Delta x.
\label{diff_geom}
\end{equation}
We also notice that
\begin{equation}
n^i n^j \nabla_i \nabla_j = (h^{i j} - \prescript{\partial}{}h^{i j}) \nabla_i \nabla_j = \nabla^2 - D^2,
\label{surface_geom}
\end{equation}
where~$\prescript{\partial}{}h^{i j}$ is the induced metric on the boundary~$\partial \Sigma^0$ and~$D$ its associated connection. If we replace the results~(\ref{equiv_bound}-\ref{surface_geom}) in the surface integral in~(\ref{beta_hat_perturb_int_2}), we obtain
\begin{multline}
\int_{\partial \Sigma^0} \rmd S^0\ \Psi^0_m \vec{n} \cdot \nabla \Delta \Psi_n \equiv \int_{\partial \Sigma^0} \rmd S^0 (\nabla^i \Delta x) (\nabla_i \Psi^0_n) \Psi^0_m \\
- \int_{\partial \Sigma^0} \rmd S^0\ \Delta x (\nabla^2 \Psi^0_n) \Psi^0_m + \int_{\partial \Sigma^0} \rmd S^0\ \Delta x (D^2 \Psi^0_n) \Psi^0_m \\
- \omega^0_n (\omega^0_n + \omega^0_m) \int_{\partial \Sigma^0} \rmd S^0\ \Delta x\ \Psi^0_n \Psi^0_m.
\label{surface_integral_split}
\end{multline}

Let us work out the first integral on the r.h.s. We can use the boundary condition~(\ref{boundary_zeroth}) to exchange the full connection~$\nabla$ and the induced connection~$D$ any time the normal component vanishes. With this in mind, we have:
\begin{align}
\int_{\partial \Sigma^0} \rmd S^0 & (\nabla^i \Delta x) (\nabla_i \Psi^0_n) \Psi^0_m \nonumber\\
= &\ \int_{\partial \Sigma^0} \rmd S^0 (D^i \Delta x) (D_i \Psi^0_n) \Psi^0_m \nonumber\\
= &\ - \int_{\partial \Sigma^0} \rmd S^0\ \Delta x (D^2 \Psi^0_n) \Psi^0_m \nonumber\\
&\ - \int_{\partial \Sigma^0} \rmd S^0 \Delta x (D_i \Psi^0_n) (D^i \Psi^0_m) \nonumber\\
= &\ - \int_{\partial \Sigma^0} \rmd S^0\ \Delta x (D^2 \Psi^0_n) \Psi^0_m \nonumber\\
&\ - \int_{\partial \Sigma^0} \rmd S^0 \Delta x (\nabla_i \Psi^0_n) (\nabla^i \Psi^0_m),
\label{surface_workout}
\end{align}
where in the second step we have used the divergence theorem and the fact that~$\partial \Sigma^0$ has no boundary.

The second integral on the r.h.s.\ of~(\ref{surface_integral_split}) can be simplified with the definition of the operator~$\hat{\mathscr{O}}^0$ and equation~(\ref{op_zeroth}), obtaining
\begin{multline}
- \int_{\partial \Sigma^0} \rmd S^0\ \Delta x (\nabla^2 \Psi^0_n) \Psi^0_m = \\
\int_{\partial \Sigma^0} \rmd S^0\ \Delta x [(\omega^0_n)^2 - \xi R^{h^0} - m^2] \Psi^0_n \Psi^0_m.
\label{surface_other}
\end{multline}
Replacing~(\ref{surface_integral_split}-\ref{surface_other}) in~(\ref{beta_hat_perturb_int_2}) we obtain~(\ref{beta_hat}), completing the proof.

The other entries of~$\Delta K (\tilde{t})$ are computed in a similar way. The only relevant difference is that for~$\Delta \hat{\alpha}_{nm} (\tilde{t})$ the equivalence relation for resonances reads
\begin{equation}
\frac{\rmd X (\tilde{t})}{\rmd \tilde{t}} \equiv \rmi (\omega^0_n - \omega^0_m) X(\tilde{t}).
\label{equiv_res_alpha}
\end{equation}
Notice in particular that this equivalence relation implies that any term in~$\Delta \hat{\alpha}_{nn} (\tilde{t})$ is equivalent to zero, and thus that these diagonal elements [but not the~$\Delta \hat{\beta}_{nn} (\tilde{t})$] are irrelevant.

\section{Derivation of the expression for~$\Delta J^\pm_{nm}$}\label{proof_J}

We obtain~(\ref{value_J}) indirectly by using the first order in~$\varepsilon$ of equation~(\ref{compare_sp}). If we pick the~$++$ sign prescripts in that equation, using~(\ref{boundary_zeroth}) and~(\ref{psi_zero_norm}) the first order in~$\varepsilon$ reads
\begin{multline}
\omega^0_n (\omega^0_n + \omega^0_m) \left[ \Delta J^+_{nm} + \int_{\Sigma^0} \rmd V^0\ \prescript{+}{}\Delta \Psi_n \Psi^0_m \right] \\
+ (\prescript{+}{}\Delta \omega_n) \delta_{nm} + \int_{\partial \Sigma^0} \rmd S^0\ \Psi^0_m \left[ \Delta x\ \vec{n} \cdot \nabla_{h^0} \left(\vec{n} \cdot \nabla_{h^0} \Psi^0_n \right) \right. \\
\left. + \vec{n} \cdot \nabla_{h^0} \prescript{+}{}\Delta \Psi_n \right] = 0,
\label{Jpp_ex}
\end{multline}
where the first term of the surface integral comes from the contribution due to the displacement of the surface~$\partial \Sigma_{\tilde{t}}$ with~$\varepsilon$. We can simplify the integrand of the surface integral using~(\ref{boundary_first}). Doing the same procedure with all of the sign prescripts, we obtain
\begin{align}
\omega^0_n (\omega^0_n + \omega^0_m) \left[ \Delta J^+_{nm} + \int_{\Sigma^0} \rmd V^0\ \prescript{+}{}\Delta \Psi_n \Psi^0_m \right] & \nonumber\\
+ (\prescript{+}{}\Delta \omega_n) \delta_{nm} - \omega^0_n \int_{\partial \Sigma^0} \rmd S^0 \left( \frac{\rmd}{\rmd \tilde{t}} \Delta x \right) \Psi^0_n \Psi^0_m & = 0, \label{Jpp} \\
\omega^0_n (\omega^0_n - \omega^0_m) \left[ \Delta J^-_{nm}+ \int_{\Sigma^0} \rmd V^0\ \prescript{+}{}\Delta \Psi_n \Psi^0_m \right] & \nonumber\\
+ \frac{1}{2}\left[(\prescript{+}{}\Delta \omega_n) - (\prescript{-}{}\Delta \omega_n)\right] \delta_{nm} & \nonumber\\
- \omega^0_n \int_{\partial \Sigma^0} \rmd S^0 \left( \frac{\rmd}{\rmd \tilde{t}} \Delta x \right) \Psi^0_n \Psi^0_m & = 0, \label{Jpm} \\
\omega^0_n (\omega^0_n - \omega^0_m) \left[ \Delta J^+_{nm} + \int_{\Sigma^0} \rmd V^0\ \prescript{-}{}\Delta \Psi_n \Psi^0_m \right] & \nonumber\\
+ \frac{1}{2}\left[(\prescript{-}{}\Delta \omega_n) - (\prescript{+}{}\Delta \omega_n)\right] \delta_{nm} & \nonumber\\
+ \omega^0_n \int_{\partial \Sigma^0} \rmd S^0 \left( \frac{\rmd}{\rmd \tilde{t}} \Delta x \right) \Psi^0_n \Psi^0_m & = 0, \label{Jmp} \\
\omega^0_n (\omega^0_n + \omega^0_m) \left[ \Delta J^-_{nm} + \int_{\Sigma^0} \rmd V^0\ \prescript{-}{}\Delta \Psi_n \Psi^0_m \right] & \nonumber\\
+ (\prescript{-}{}\Delta \omega_n) \delta_{nm} + \omega^0_n \int_{\partial \Sigma^0} \rmd S^0 \left( \frac{\rmd}{\rmd \tilde{t}} \Delta x \right) \Psi^0_n \Psi^0_m & = 0. \label{Jmm}
\end{align}
Adding~(\ref{Jpp}) and~(\ref{Jmp}) on the one side, and~(\ref{Jpm}) and~(\ref{Jmm}) on the other side, and solving for~$\Delta J^\pm_{nm}$ respectively, one obtains~(\ref{value_J}).

\section{Derivation of the differential geometry relation}\label{proof_geom}

In this Appendix we prove the relation~(\ref{diff_geom})~\cite{stefan}. Let us consider the family of surfaces~$\partial \Sigma_{\tilde{t}} (\varepsilon)$ around~$\partial \Sigma^0$ given by different values of~$\varepsilon$, and~$n^i$ the vector field normal to the family of surfaces. In an arbitrarily close neighbourhood of~$\partial \Sigma^0$ we can use the coordinate chart
\begin{equation}
(x^1 = \varepsilon, x^2, \ldots, x^N),
\label{coords}
\end{equation}
where the vector fields~$\partial_i$ for $i>1$ are tangent to the surfaces. Notice that in this coordinate chart the metric is such that $h^{1i} = 0$ for $i>1$. By definition, the quantity~$\Delta x$ is the proper length displacement per unit~$\varepsilon$. It is therefore given by
\begin{equation}
\Delta x = \| \frac{\rmd}{\rmd \varepsilon} \left( \begin{array}{c}
	x^1 \\
	x^2 \\
	\ldots \\
	x^N
\end{array} \right) \| =
\| \left( \begin{array}{c}
	1 \\
	0 \\
	\ldots \\
	0
\end{array} \right) \| =
\sqrt{h_{11}},
\label{def_dx}
\end{equation}
where~$\| \cdot \|$ is the norm given by the metric. The vector~$n^i$ is pointing in the direction~$\partial_1$ and has unit norm. Therefore, $n^i = \delta^i_1/\Delta x$. Using this last expression we can easily compute
\begin{align}
n^i \nabla_i n^j & = \frac{1}{\Delta x} \nabla_1 n^j = \frac{1}{\Delta x} ( \partial_1 n^j + \Gamma^j_{1i} n^i ) \nonumber\\
& = \frac{1}{\Delta x^2} \left( \Gamma^j_{11} - \delta^j_1 \frac{\partial_1 \Delta x}{\Delta x} \right). \label{ndn_1}
\end{align}

We have to compute the Christoffel symbols appearing in~(\ref{ndn_1}). We do so by using $h^{1i} = 0$ for $i>1$, and $h_{11} = \Delta x^2$ from~(\ref{def_dx}), finding
\begin{align}
\Gamma^j_{11} & = \frac{1}{2} h^{ij}(2 \partial_1 h_{1i} - \partial_i h_{11}) \nonumber\\
& = \frac{1}{2} \left( h^{1j} \partial_1 h_{11} - \sum_{i>1} h^{ij} \partial_i h_{11} \right) \nonumber\\
& = \delta^j_1 \frac{\partial_1 \Delta x}{\Delta x} - \Delta x \sum_{i>1} h^{ij} \partial_i \Delta x.
\label{Gammas}
\end{align}
By replacing~(\ref{Gammas}) in~(\ref{ndn_1}) we find that
\begin{align}
n^i \nabla_i n^j & = - \frac{1}{\Delta x} \sum_{i>1} h^{ij} \partial_i \Delta x \nonumber\\
& = \frac{1}{\Delta x} (n^i n^j - h^{i j}) \nabla_i \Delta x.
\label{ndn_2}
\end{align}
The last expression can be obtained from the previous one by checking independently that they coincide both for $j = 1$ and for $j > 1$. This expression is manifestly covariant and therefore valid in any coordinate chart.

\section{Dirichlet vanishing boundary conditions}\label{dirichlet_app}

The construction of the method for Dirichlet vanishing boundary conditions is analogous to the one done for Neumann vanishing boundary conditions. That is, we also construct bases of modes associated to Cauchy hypersurfaces satisfying Properties~I and~II, and then find a differential equation in time for the linear transformation between these bases, which can be interpreted as a Bogoliubov transformation when the conditions given in Property~II are met.

However, the imposition of Dirichlet boundary conditions with moving boundaries in a way that makes them compatible with the application of the method is more intricate than what it may seem on a first stage. One may think that condition~(\ref{dirichlet}) does not relate the field with its time derivative, as Neumann condition does, and therefore that a construction like the one done for the Neumann condition in this second article is not necessary, and one can proceed as in Part~I. However, when~(\ref{dirichlet}) is considered as a boundary condition constraining the possible initial conditions on a Cauchy hypersurface~$\Sigma_t$, it should be taken into account also how the time derivative of this global boundary condition may constrain the first partial time derivative of a mode at the boundary of the hypersurface. If one takes the total derivative with respect to~$t$ along the boundary of~(\ref{dirichlet}), it is easy to obtain
\begin{equation}
\partial_t \Phi (t, \vec{x})= - v_{\mathrm{B}} (t, \vec{x}) \vec{n} \cdot \nabla_{h(t)} \Phi (t, \vec{x});
\label{dirichlet_diff}
\end{equation}
which is an expression identical to the first line of~(\ref{neumann_initial}), but replacing~$v_{\mathrm{B}} \to 1/v_{\mathrm{B}}$. It is clear then that we should impose both~(\ref{dirichlet}) and~(\ref{dirichlet_diff}) to the possible initial conditions. However, imposing these two boundary conditions raises a technical difficulty for computing an useful auxiliary basis~$\{\Psi^{[\tilde{t}]}_n\}$ of initial conditions. By useful, we mean that it should be possible to linearly transform it into a basis of ``locally well-defined frequency'' modes (in the regions where those can be defined), as we did in the case of Neumann boundary conditions with the transformation~(\ref{psi_to_phi}). In order for this transformation to provide ``locally well-defined frequency'' modes, one crucial ingredient is the proportionality relation in~(\ref{first_derivative}). But this proportionality relation, together with conditions~(\ref{dirichlet}) and~(\ref{dirichlet_diff}), would imply that both~$\Psi^{[\tilde{t}]}_n(\tilde{t})$ and~$\vec{n} \cdot \nabla_{h(\tilde{t})} \Psi^{[\tilde{t}]}_n(\tilde{t})$ vanish at the boundary (except when~$v_{\mathrm{B}} = 0$). If we wished now to find the set of initial conditions as solutions to an eigenvalue problem in~$\Sigma_{\tilde{t}}$ with an elliptic operator, in general we would not find non-zero solutions.

How do we get out of this blind alley? When considering how the boundary conditions constrain the initial conditions, we renounce to require~(\ref{dirichlet}) and keep only condition~(\ref{dirichlet_diff}), \emph{except} for the regions of the boundary which remain static ($v_{\mathrm{B}} = 0$), in which case we impose both conditions. This decision might look not legitimated, since after all we will construct modes that, in general, do not satisfy Dirichlet vanishing boundary conditions. However, a careful discussion shows that the construction obtained is valid when the method is used to compute the evolution between regions for which a physically valid Fock quantisation can be done, as described in Subsect.~\ref{physical}. In particular, notice that in those regions the boundaries must remain static (although in the time between the regions they may of course not). Therefore, when constructing the initial conditions of the modes associated to those regions, we \emph{do} also impose their initial conditions to vanish at the boundary.

With that in mind, let us justify why we can leave the condition~(\ref{dirichlet}) only for the static regions of the boundary. We notice that condition~(\ref{dirichlet_diff}) (which is satisfied by all modes) was obtained by taking the total time derivative of~(\ref{dirichlet}) along the boundary. Therefore, the fulfilment of this condition guarantees that the value at the boundary is \emph{preserved} in time. That is, the bases of solutions that we obtain expand the space of solutions to the Klein-Gordon equation for which the evaluation of the field at the boundary remains constant (along the direction of the projection of~$\partial_t$ over the boundary). If we then use the method to compute the evolution of a mode which vanishes at the intersection of some Cauchy hypersurface with the boundary, we can be sure that this zero value is preserved along the whole spacetime boundary. This means that, even if for expanding this mode at a different time we use a basis of modes which, individually, do not vanish at the boundary, their linear combination given by the expansion necessarily vanishes.

In summary: We are, strictly speaking, not constructing the method \emph{directly} for Dirichlet vanishing boundary conditions. Rather, we are constructing the method for Dirichlet ``time-preserved and vanishing-when-static'' boundary conditions. But, at the same time, we are making sure to use the method \emph{only} with bases of modes that vanish at the boundary at \emph{some} time, and therefore at \emph{every} time, thus fulfilling Dirichlet vanishing boundary conditions, as desired.

Once the discussion above has been done, the remaining development of the method follows in a completely analogous way to the one done for Neumann boundary conditions. Therefore, we do not reproduce all the calculations and discussions in detail. We rather list all the objects and formulas in the article, excluding the examples and the Appendices, that change when considering Dirichlet boundary conditions. Any formula that is not listed below can be used with both boundary conditions.

In Subsect.~\ref{objects} the subspace of initial conditions $\Gamma_t \subset [C^{\infty} (\Sigma_t) \cap L^2 (\Sigma_t)]^{\oplus 2}$ is the restriction of the full space of pairs of functions to initial conditions $(\Phi, \partial_t \Phi)|_{\Sigma_t}$ satisfying
\begin{equation}
\left\{
\begin{array}{l}
	\partial_t \Phi (t, \vec{x}) = - v_{\mathrm{B}}(t, \vec{x}) \vec{n} \cdot \nabla_{h(t)} \Phi (t, \vec{x}), \\
	\text{and}\ \Phi (t, \vec{x}) = 0\ \text{if}\ v_{\mathrm{B}}(t, \vec{x}) = 0;
\end{array}
\right.
\label{dirichlet_initial}
\end{equation}
where~$\vec{x} \in \partial \Sigma_t$ [instead of~(\ref{neumann_initial})]. That is, pairs of initial conditions satisfying~(\ref{dirichlet_diff}) always and~(\ref{dirichlet}) only when the boundary is static, as we advanced.

In Subsect.~\ref{bases} equation~(\ref{neumann_moving_modes_boundary}) must be replaced by
\begin{equation}
\omega^{[\tilde{t}]}_n \Psi^{[\tilde{t}]}_n (\tilde{t}, \vec{x}) = - v_{\mathrm{B}} (\tilde{t}, \vec{x}) \vec{n} \cdot \nabla_{h(\tilde{t})} \Psi^{[\tilde{t}]}_n(\tilde{t}, \vec{x}), \quad \vec{x} \in \partial \Sigma_{\tilde{t}}.
\label{dirichlet_moving_modes_boundary}
\end{equation}
Notice that, since $\omega^{[\tilde{t}]}_n \neq 0$ (see \ref{no_zeros}), when $v_{\mathrm{B}} = 0$ this equation is also imposing the second condition in~(\ref{dirichlet_initial}).

In Subsect.~\ref{sec_linear_transf} the expression for~$\hat{V}^{\pm \hat{\pm}}_{nm}$ in~(\ref{V_hat}) is slightly changed (only the surface integral changes\footnote{The surface integral in~(\ref{V_hat}) had been simplified using~(\ref{neumann_moving_modes_boundary}). The expression given in~(\ref{V_hat_dirichlet}) is actually valid for both boundary conditions. }):
\begin{multline}
\hat{V}^{\pm \hat{\pm}}_{nm} = -(\prescript{\hat{\pm}}{}\omega_m^{[\tilde{t}]})\delta_{nm}\delta_{\pm \hat{\pm}} \\
\hat{\pm} \left\{ \left[(\prescript{\pm}{}\omega_n^{[\tilde{t}]})+(\prescript{\hat{\pm}}{}\omega_m^{[\tilde{t}]})\right]\int_{\Sigma_{\tilde{t}}}\rmd V_{\tilde{t}} \left[\frac{\rmd}{\rmd \tilde{t}}\prescript{\pm}{}\Psi_n^{[\tilde{t}]}(\tilde{t}) \right] \prescript{\hat{\pm}}{}\Psi_m^{[\tilde{t}]}(\tilde{t}) \right. \\
\left.+\left[2 (\prescript{\pm}{}\omega_n^{[\tilde{t}]} )^2+\frac{\rmd\prescript{\pm}{}\omega_n^{[\tilde{t}]}}{\rmd \tilde{t}} - F(\tilde{t})\right]\int_{\Sigma_{\tilde{t}}}\rmd V_{\tilde{t}}\prescript{\pm}{}\Psi_n^{[\tilde{t}]}(\tilde{t})\prescript{\hat{\pm}}{}\Psi_m^{[\tilde{t}]}(\tilde{t}) \right. \\
\left.+\int_{\Sigma_{\tilde{t}}}\rmd V_{\tilde{t}}\prescript{\pm}{}\Psi_n^{[\tilde{t}]}(\tilde{t})\left[\prescript{\pm}{}\omega_n^{[\tilde{t}]}q(\tilde{t})+\xi\bar{R}(\tilde{t})\right]\prescript{\hat{\pm}}{}\Psi_m^{[\tilde{t}]}(\tilde{t}) \right. \\
\left. + \frac{1}{\prescript{\hat{\pm}}{}\omega_m^{[\tilde{t}]}} \int_{\partial \Sigma_{\tilde{t}}}\rmd S_{\tilde{t}}\ \left[\frac{\rmd}{\rmd \tilde{t}}\prescript{\pm}{}\Psi_n^{[\tilde{t}]}(\tilde{t}) \right] \vec{n} \cdot \nabla_{h(\tilde{t})} \prescript{\hat{\pm}}{}\Psi^{[\tilde{t}]}_m(\tilde{t}) \right\}.
\label{V_hat_dirichlet}
\end{multline}

Finally, in Sect.~\ref{res} equation~(\ref{boundary_zeroth}) becomes
\begin{equation}
\Psi^0_n(\vec{x}) = 0, \quad \vec{x} \in \partial \Sigma^0;
\label{boundary_zeroth_dirichlet}
\end{equation}
and the entries of the matrix $\Delta K(\tilde{t})$ in~(\ref{perturbation_matrix}), given by~(\ref{alpha_hat}) and~(\ref{beta_hat}), must be replaced by
\begin{align}
\Delta &\hat{\alpha}_{n m} (\tilde{t}) \equiv \ \rmi \int_{\Sigma^0} \rmd V^0\ [\prescript{-}{m}{\hat{\Delta}}(\tilde{t}) \Psi^0_n] \Psi^0_m \nonumber \\
& - \rmi \int_{\partial \Sigma_{0}} \rmd S^0\ \Delta x(\tilde{t}) \left(\vec{n}\cdot\nabla_{h^0}\Psi_n^0\right)\left(\vec{n}\cdot\nabla_{h^0}\Psi_m^0\right), \label{ssd_alpha_hat_moving} \\
\Delta &\hat{\beta}_{n m} (\tilde{t}) \equiv \ -\rmi \int_{\Sigma^0} \rmd V^0\ [\prescript{+}{m}{\hat{\Delta}}(\tilde{t}) \Psi^0_n] \Psi^0_m \nonumber \\
& + \rmi \int_{\partial \Sigma_{0}} \rmd S^0\ \Delta x(\tilde{t}) \left(\vec{n}\cdot\nabla_{h^0}\Psi_n^0\right)\left(\vec{n}\cdot\nabla_{h^0}\Psi_m^0\right). \label{ssd_beta_hat_moving}
\end{align}

\section{Proof that $\omega^{[\tilde{t}]}_n \neq 0$}\label{no_zeros}

Let us consider a solution of~(\ref{eigenvalues_good}) for which $\omega^{[\tilde{t}]}_n = 0$. Then there exists a function~$\Psi^{[\tilde{t}]}_n(\tilde{t})$ satisfying~(\ref{neumann_moving_modes}) and~(\ref{neumann_moving_modes_boundary}) with~$\omega^{[\tilde{t}]}_n = 0$. But~(\ref{neumann_moving_modes_boundary}) with~$\omega^{[\tilde{t}]}_n = 0$ are simply Neumann vanishing boundary conditions for the \emph{spatial} function~$\Psi^{[\tilde{t}]}_n(\tilde{t})$, and the operator~$\hat{\mathscr{O}}(\tilde{t})$ is clearly positive definite for functions satisfying such boundary conditions. Since, according to~(\ref{neumann_moving_modes}), $\Psi^{[\tilde{t}]}_n(\tilde{t})$ is an eigenfunction of this operator with eigenvalue~$(\omega^{[\tilde{t}]}_n)^2$, we must have~$(\omega^{[\tilde{t}]}_n)^2 > 0$, which is a contradiction. Therefore~$\omega^{[\tilde{t}]}_n \neq 0$.

In the case of Dirichlet boundary conditions, the proof that $\omega^{[\tilde{t}]}_n \neq 0$ is slightly more subtle. We also start by considering a solution of~(\ref{eigenvalues_good}) for which $\omega^{[\tilde{t}]}_n = 0$. Then there exists a function~$\Psi^{[\tilde{t}]}_n(\tilde{t})$ satisfying~(\ref{neumann_moving_modes}) and~(\ref{dirichlet_moving_modes_boundary}) with~$\omega^{[\tilde{t}]}_n = 0$. But~(\ref{dirichlet_moving_modes_boundary}) with~$\omega^{[\tilde{t}]}_n = 0$ implies that the spatial function~$\Psi^{[\tilde{t}]}_n(\tilde{t})$ satisfies Neumann vanishing boundary conditions at least in the regions of the boundary where $v_{\mathrm{B}} \neq 0$. But in the regions of the boundary where $v_{\mathrm{B}} = 0$ we know that the second line of~(\ref{dirichlet_initial}) must be imposed, and therefore the function~$\Psi^{[\tilde{t}]}_n(\tilde{t})$ must satisfy Dirichlet vanishing boundary conditions there.\footnote{Notice that~(\ref{dirichlet_moving_modes_boundary}) implies the second line of~(\ref{dirichlet_initial}) only \emph{after} having already found that $\omega^{[\tilde{t}]}_n \neq 0$. Since here we are assuming the opposite, we must impose the second line of~(\ref{dirichlet_initial}) explicitly.} Therefore, the function ~$\Psi^{[\tilde{t}]}_n(\tilde{t})$ satisfies mixed vanishing boundary conditions. But the operator~$\hat{\mathscr{O}}(\tilde{t})$ is clearly positive definite for functions satisfying such boundary conditions. Then, again according to~(\ref{neumann_moving_modes}) we must have~$(\omega^{[\tilde{t}]}_n)^2 > 0$, which is a contradiction. Therefore~$\omega^{[\tilde{t}]}_n \neq 0$.

\section{Summary of formulae for the application of the method}\label{recipes}

In this Appendix, we provide Tables~\ref{summary} and~\ref{summary2} with all the formulae necessary in order to apply the method to a concrete problem. As we indicated in the examples in Subsect.\ \ref{DCE} and \ref{gw}, once the expressions for the method have been found, there is no need to consider the explicit evolution in time of the modes constructed any more. Therefore, the notation for the ``time label'' of the different modes can be simplified replacing~$\tilde{t} \to t$. In the Tables we use this simplification.

\begin{table*}
\caption{Summary of formulae for the application of the method (part~1)}
\label{summary}
\begin{tabular*}{\textwidth}{@{}lc}
\hline
\textbf{Computation of the auxiliary modes and eigenvalues}\\
\hline
(Pseudo-)eigenvalue equation (\ref{neumann_moving_modes})&
$\begin{aligned}
\hat{\mathscr{O}}(t) \Psi^{[t]}_n (t) & = (\omega^{[t]}_n)^2 \Psi^{[t]}_n (t), \quad \Psi^{[t]}_n (t)\ \text{real};
\end{aligned}$
\vspace{0.2cm}
\\
with the operator (\ref{operator}, \ref{def_F})&
$\begin{aligned}
\hat{\mathscr{O}}(t) & = - \nabla_{h(t)}^2 + \xi R^h(t) + m^2 + F(t), \\
F(t) & =
\left\{
\begin{array}{l}
	0 \quad \text{if}\ \xi R^h (t, \vec{x}) + m^2 > 0\ \text{a.e.\ in}\ \Sigma_t,\\
	-\text{ess\ inf} \{\xi R^h (t, \vec{x}) + m^2, (t,\vec{x}) \in \Sigma_t\} + \epsilon \quad \text{i.o.c.};
\end{array}
\right.
\end{aligned}$
\vspace{0.2cm}
\\
boundary conditions (for $\vec{x} \in \partial \Sigma_{t}$) (\ref{neumann_moving_modes_boundary}, \ref{dirichlet_moving_modes_boundary})&
$\begin{aligned}
\vec{n} \cdot \nabla_{h(t)} \Psi^{[t]}_n(t) & = - \omega^{[t]}_n v_{\mathrm{B}}(t) \Psi^{[t]}_n (t) & \text{(Neumann)}, \\
\omega^{[t]}_n \Psi^{[t]}_n (t) & = - v_{\mathrm{B}}(t) \vec{n} \cdot \nabla_{h(t)} \Psi^{[t]}_n(t) & \text{(Dirichlet)}; \\
\end{aligned}$
\vspace{0.2cm}
\\
and orthonormalisation condition (\ref{norma_casera}, \ref{first_derivative}, \ref{scalar_product_ini})&
$\begin{aligned}
& \int_{\Sigma_t} \rmd V_t\ [\xi R^h (t) + m^2 + F(t) + \omega^{[t]}_n \omega^{[t]}_m] \Psi^{[t]}_n (t) \Psi^{[t]}_m (t) \\
& + \int_{\Sigma_t} \rmd V_t\ [\nabla_{h(t)} \Psi^{[t]}_n (t)] \cdot [\nabla_{h(t)} \Psi^{[t]}_m (t)] = |\omega^{[t]}_n| \delta_{n m}.
\end{aligned}$\\
\hline
\textbf{Time-dependent linear transformation}\\
\hline
Differential equation~(\ref{differential_equation_moving_U})&
$\begin{aligned}
\frac{\rmd}{\rmd t} U(t, t_0) = M \hat{V}(t) M^* U(t, t_0);
\end{aligned}$
\vspace{0.2cm}
\\
with (\ref{matrix_M}, \ref{def_Vhat}, \ref{V_hat_dirichlet}, \ref{V_hat}, \ref{change_factor}, \ref{bar_scalar})&
$\begin{aligned}
& M = \frac{1}{2}
\left( \begin{array}{cc}
	(1-\rmi) I & (1+\rmi) I \\
	(1+\rmi) I & (1-\rmi) I
\end{array} \right), \quad
\hat{V}(t) =
\left(
\begin{array}{cc}
	\hat{V}^{++} & \hat{V}^{+-} \\
	\hat{V}^{-+} & \hat{V}^{--}
\end{array}
\right),\\
& \hat{V}^{\pm \hat{\pm}}_{nm} = -(\prescript{\hat{\pm}}{}\omega_m^{[t]})\delta_{nm}\delta_{\pm \hat{\pm}} \\
&\ \hat{\pm} \left\{ \left[(\prescript{\pm}{}\omega_n^{[t]})+(\prescript{\hat{\pm}}{}\omega_m^{[t]})\right]\int_{\Sigma_{t}}\rmd V_{t} \left[\frac{\rmd}{\rmd t}\prescript{\pm}{}\Psi_n^{[t]}(t) \right] \prescript{\hat{\pm}}{}\Psi_m^{[t]}(t) \right. \\
&\ \left.+\left[2 (\prescript{\pm}{}\omega_n^{[t]} )^2+\frac{\rmd\prescript{\pm}{}\omega_n^{[t]}}{\rmd t} - F(t)\right]\int_{\Sigma_{t}}\rmd V_{t}\prescript{\pm}{}\Psi_n^{[t]}(t)\prescript{\hat{\pm}}{}\Psi_m^{[t]}(t) \right. \\
&\ \left.+\int_{\Sigma_{t}}\rmd V_{t}\prescript{\pm}{}\Psi_n^{[t]}(t)\left[\prescript{\pm}{}\omega_n^{[t]}q(t)+\xi\bar{R}(t)\right]\prescript{\hat{\pm}}{}\Psi_m^{[t]}(t) + \mathrm{(SI)} \right\};\\
& \mathrm{(SI)} := \left\{
\begin{array}{l}
(\prescript{\hat{\pm}}{}\omega_m^{[t]})^{-1} \int_{\partial \Sigma_{t}}\rmd S_{t}\ \left[\frac{\rmd}{\rmd t}\prescript{\pm}{}\Psi_n^{[t]}(t) \right] \vec{n} \cdot \nabla_{h(t)} \prescript{\hat{\pm}}{}\Psi^{[t]}_m(t)\\
\text{(Dirichlet-Neumann)},\\
	- \int_{\partial \Sigma_{t}}\rmd S_{t}\ v_{\mathrm{B}}(t) \left[\frac{\rmd}{\rmd t}\prescript{\pm}{}\Psi_n^{[t]}(t) \right] \prescript{\hat{\pm}}{}\Psi_m^{[t]}(t) \\
\text{(Neumann)};\\
\end{array}
\right.\\
& q(t) = \partial_t \log \sqrt{h(t)},\\
& \bar{R}(t) = 2 \partial_t q(t) + q(t)^2 - [\partial_t h^{i j} (t)] [\partial_t h_{i j} (t)]/4.
\end{aligned}$
\vspace{0.2cm}
\\
Formal solution (\ref{U_moving_solution})&
$\begin{aligned}
U(t_{\mathrm{f}}, t_0) = \mathscr{T} \exp \left[ \int_{t_0}^{t_{\mathrm{f}}} \rmd t\ M \hat{V}(t) M^* \right].
\end{aligned}$\\
\hline
\end{tabular*}
\end{table*}

\newpage

\begin{table*}
\caption{Summary of formulae for the application of the method (part~2)}
\label{summary2}
\begin{tabular*}{\textwidth}{@{}lc}
\hline
\textbf{Perturbative regime}\\
\hline
Quantities~$\Delta \hat{\alpha}$ and~$\Delta \hat{\beta}$\\
\hline
Neumann boundary conditions (\ref{alpha_hat}, \ref{beta_hat})&
$\begin{aligned}
\Delta \hat{\alpha}_{n m} (t) \equiv &\ \rmi \int_{\Sigma^0} \rmd V^0\ [\prescript{-}{m}{\hat{\Delta}}(t) \Psi^0_n] \Psi^0_m \\
& + \rmi \int_{\partial \Sigma^0} \rmd S^0\ \Delta x(t) \Big[ (\nabla_{h^0} \Psi^0_n) \cdot (\nabla_{h^0} \Psi^0_m) \\
& + (\xi R^{h^0} + m^2 - \omega^0_n \omega^0_m) \Psi^0_n \Psi^0_m \Big], \\
\Delta \hat{\beta}_{n m} (t) \equiv & - \rmi \int_{\Sigma^0} \rmd V^0\ [\prescript{+}{m}{\hat{\Delta}}(t) \Psi^0_n] \Psi^0_m \\
& - \rmi \int_{\partial \Sigma^0} \rmd S^0\ \Delta x(t) \Big[ (\nabla_{h^0} \Psi^0_n) \cdot (\nabla_{h^0} \Psi^0_m) \\
& + (\xi R^{h^0} + m^2 + \omega^0_n \omega^0_m) \Psi^0_n \Psi^0_m \Big];
\end{aligned}$
\vspace{0.2cm}
\\
Dirichlet boundary conditions (\ref{ssd_alpha_hat_moving}, \ref{ssd_beta_hat_moving})&
$\begin{aligned}
\Delta \hat{\alpha}_{n m} (t) \equiv &\ \rmi \int_{\Sigma^0} \rmd V^0\ [\prescript{-}{m}{\hat{\Delta}}(t) \Psi^0_n ] \Psi^0_m \\
& - \rmi \int_{\partial \Sigma_{0}} \rmd S^0\ \Delta x(t) \left(\vec{n}\cdot\nabla_{h^0}\Psi_n^0\right)\left(\vec{n}\cdot\nabla_{h^0}\Psi_m^0\right), \\
\Delta \hat{\beta}_{n m} (t) \equiv & - \rmi \int_{\Sigma^0} \rmd V^0\ [\prescript{+}{m}{\hat{\Delta}}(t) \Psi^0_n ] \Psi^0_m \\
& + \rmi \int_{\partial \Sigma_{0}} \rmd S^0\ \Delta x(t) \left(\vec{n}\cdot\nabla_{h^0}\Psi_n^0\right)\left(\vec{n}\cdot\nabla_{h^0}\Psi_m^0\right);
\end{aligned}$
\vspace{0.2cm}
\\
with the static modes (\ref{op_zeroth})&
$\begin{aligned}
\hat{\mathscr{O}}^0 \Psi^0_n = (\omega^0_n)^2 \Psi^0_n, \quad \Psi^0_n\ \text{real};
\end{aligned}$
\vspace{0.2cm}
\\
with boundary conditions (for $\vec{x} \in \partial \Sigma^0$) (\ref{boundary_zeroth}, \ref{boundary_zeroth_dirichlet})&
$\begin{aligned}
\vec{n} \cdot \nabla_{h^0} \Psi^0_n(\vec{x}) & = 0 & \text{(Neumann)},\\
\Psi^0_n(\vec{x}) & = 0 & \text{(Dirichlet)};
\end{aligned}$
\vspace{0.2cm}
\\
and orthonormalisation condition (\ref{psi_zero_norm})&
$\begin{aligned}
\int_{\Sigma^0} \rmd V^0\ \Psi^0_n \Psi^0_m = \frac{\delta_{n m}}{2 \omega^0_n};
\end{aligned}$
\vspace{0.2cm}
\\
and with the operators (\ref{superoperator})&
$\begin{aligned}
\prescript{\pm}{m}{\hat{\Delta}}(t) \Psi^0_n = \big[ \Delta \hat{\mathscr{O}} (t) + \omega^0_n (\omega^0_n \pm \omega^0_m) \Delta r(t) + \xi \Delta \bar{R}(t) \big] \Psi^0_n,
\end{aligned}$
\vspace{0.2cm}
\\
with (\ref{def_dr})&
$\begin{aligned}
\Delta r (t) = \partial_\varepsilon \log h(t) |_{\varepsilon = 0} / 2,
\end{aligned}$\\
and &
$\begin{aligned}
\Delta F (t) = 0.
\end{aligned}$\\
\hline
Bogoliubov coefficients\\
\hline
Explicit time evolution (resonances) (\ref{perturbation_alpha}, \ref{perturbation_beta})&
$\begin{aligned}
\alpha_{n n} (t_{\mathrm{f}}, t_0) & \approx 1;\\
\alpha_{n m} (t_{\mathrm{f}}, t_0) & \approx \varepsilon \int_{t_0}^{t_{\mathrm{f}}} \rmd t\ \rme^{-\rmi (\omega^0_n - \omega^0_m) t} \Delta \hat{\alpha}_{n m} (t), \quad n \neq m;\\
\beta_{n m} (t_{\mathrm{f}}, t_0) & \approx \varepsilon \int_{t_0}^{t_{\mathrm{f}}} \rmd t\ \rme^{-\rmi (\omega^0_n + \omega^0_m) t} \Delta \hat{\beta}_{n m} (t).
\end{aligned}$
\vspace{0.2cm}
\\
Asymptotic values using Fourier transforms (\ref{fourier_alpha}, \ref{fourier_beta})&
$\begin{aligned}
\alpha_{n n} (-\infty,\infty) & \approx 1;\\
\alpha_{n m} (-\infty,\infty) & \approx \varepsilon \sqrt{2 \pi}\ \mathscr{F} [\Delta \hat{\alpha}_{n m}] (\omega^0_n - \omega^0_m), \quad n \neq m;\\
\beta_{n m} (-\infty,\infty) & \approx \varepsilon \sqrt{2 \pi}\ \mathscr{F} [\Delta \hat{\beta}_{n m}] (\omega^0_n + \omega^0_m).
\end{aligned}$\\
\hline
\end{tabular*}
\end{table*}

\bibliographystyle{spphys} 
\bibliography{ContBogos} 

\end{document}